\documentclass[acmsmall,screen,authorversion]{acmart}

\AtBeginDocument{%
	\providecommand\BibTeX{{%
			\normalfont B\kern-0.5em{\scshape i\kern-0.25em b}\kern-0.8em\TeX}}}

\usepackage{booktabs} 

\usepackage{graphicx}
\usepackage{color}
\usepackage{listings}
\usepackage{pifont}
\usepackage{array}
\usepackage{xr}
\usepackage{multirow}
\usepackage{longtable}
\usepackage{import}
\usepackage{bibunits}
\usepackage{url}
\usepackage{enumerate}
\usepackage{enumitem}
\usepackage[most]{tcolorbox}

\usepackage{lineno}
\usepackage{hyperref}
\usepackage{comment}
\usepackage{rotating}
\usepackage[normalem]{ulem}
\usepackage{makecell}
\usepackage{balance}
\usepackage[most]{tcolorbox}

\usepackage[ruled]{algorithm2e} 

\SetAlFnt{\small}
\SetAlCapFnt{\small}
\SetAlCapNameFnt{\small}
\SetAlCapHSkip{0pt}
\IncMargin{-\parindent}

\usepackage[colorinlistoftodos]{todonotes}
\usepackage{verbatim}

\usepackage{geometry}

\acmJournal{TOCE}

\setcopyright{acmcopyright}
\copyrightyear{2021}
\acmYear{2021}

\begin{document}
\title[Empiricism in CER]{A Systematic Literature Review of Empiricism and Norms of Reporting in Computing Education Research Literature}  

\author{Sarah Heckman}
\orcid{0000-0003-4351-8611}
\affiliation{%
  \institution{North Carolina State University}
}
\email{sarah_heckman@ncsu.edu}

\author{Jeffrey C. Carver}
\orcid{0000-0002-7824-9151}
\affiliation{%
  \institution{University of Alabama}
}
\email{carver@cs.ua.edu}

\author{Mark Sherriff}
\orcid{0000-0002-1745-205X}
\affiliation{%
  \institution{University of Virginia}
}
\email{sherriff@virginia.edu}

\author{Ahmed Al-Zubidy}
\orcid{0000-0003-4305-2153}
\affiliation{%
  \institution{University of Alabama}
}
\email{aalzubidy@crimson.ua.edu}

\newcommand{\ot}[1]{{\color{blue}}}
\newcommand{\jnote}[1]{{\color{blue}JEFF: #1}}
\newcommand{\mnote}[1]{{\color{red}MARK: #1}}
\newcommand{\editorial}[1]{}
\newcommand{\editorialdone}[1]{}

\begin{abstract}

\textbf{Context.} Computing Education Research (CER) is critical to help the computing education community and policy makers support the increasing population of students who need to learn computing skills for future careers. For a community to systematically advance knowledge about a topic, the members must be able to understand published work thoroughly enough to perform replications, conduct meta-analyses, and build theories. There is a need to understand whether published research allows the CER community to systematically advance knowledge and build theories.

\textbf{Objectives.}  The goal of this study is \textit{to characterize the reporting of empiricism in Computing Education Research literature by identifying whether publications include content necessary for researchers to perform replications, meta-analyses, and theory building.} We answer three research questions related to this goal: RQ1) What percentage of papers in CER venues have some form of empirical evaluation? RQ2) Of the papers that have empirical evaluation, what are the characteristics of the empirical evaluation? RQ3) Of the papers that have empirical evaluation, do they follow norms (both for inclusion and for labeling of information needed for replication, meta-analysis, and, eventually, theory-building) for reporting empirical work? 

\textbf{Methods.} We conducted a systematic literature review of the 2014 and 2015 proceedings or issues of five CER venues: \textit{Technical Symposium on Computer Science Education} (SIGCSE TS), \textit{International Symposium on Computing Education Research} (ICER), \textit{Conference on Innovation and Technology in Computer Science Education} (ITiCSE), \textit{ACM Transactions on Computing Education} (TOCE), and \textit{Computer Science Education} (CSE). We developed and applied the \textit{CER Empiricism Assessment Rubric} to the 427 papers accepted and published at these venues over 2014 and 2015. Two people evaluated each paper using the \textit{Base Rubric} for characterizing the paper. An individual person applied the other rubrics to characterize the norms of reporting, as appropriate for the paper type. Any discrepancies or questions were discussed between multiple reviewers to resolve.

\textbf{Results.} We found that over 80\% of papers accepted across all five venues had some form of empirical evaluation. Quantitative evaluation methods were the most frequently reported.  Papers most frequently reported results on interventions around pedagogical techniques, curriculum, community, or tools. There was a split in papers that had some type of comparison between an intervention and some other data set or baseline.  Most papers reported related work, following the expectations for doing so in the SIGCSE and CER community.  However, many papers were lacking properly reported research objectives, goals, research questions, or hypotheses; description of participants; study design; data collection; and threats to validity.  These results align with prior surveys of the CER literature.

\textbf{Conclusions.} CER authors are contributing empirical results to the literature; however, not all norms for reporting are met. We encourage authors to provide clear, labeled details about their work so that readers can use the study methodologies and results for replications and meta-analyses. As our community grows, our reporting of CER should mature to help establish computing education theory to support the next generation of computing learners.

\end{abstract}

%
%


%
%


\maketitle

\renewcommand{\shortauthors}{Heckman, Carver, Sherriff, and Al-Zubidy}

\section{Introduction}
\label{introduction}

From 2009 to 2015, the number of bachelor's degrees in Computer Science increased by 74 percent, while overall growth across all fields of study only rose by 16 percent~\cite{NAP24926}.  Taulbee's 2019~\cite{taulbee2019} report shows an increase from a recent minimum of under 10,000 bachelor's degrees conferred in 2009 to over 30,000 degrees in 2019 (see Figure 1B~\cite{taulbee2019}).
It is undeniable that there has been a rapid increase in interest in computer science courses at higher-education institutions.  
We see similar demand in K-12 with CSforAll\footnote{See https://www.csforall.org/ for more details.} initiatives.
Educators have struggled with how to cope with new and exacerbated challenges in computing education - ranging from how to scale up to handle new enrollments to which teaching techniques are best during this time of growth.  The COVID-19 pandemic has added additional investigation into the challenges of online instruction.  

At the same time, attendance at the ACM SIGCSE Technical Symposium has continued to increase year over year, coinciding with a similar increase in the number of submissions~\cite{Hawthorne:2019SIGCSEReport,Zhang:2020SIGCSETSRecap}.  
Computing faculty are eagerly trying new teaching methodologies to address the challenges they face with growth in enrollment and with ever-emerging technologies.  
Other computing education research (CER) venues are also seeing increases.  
The SIGCSE Board~\footnote{We use SIGCSE to represent the ACM Special Interest Group in Computer Science Education as a community of computing education researchers and practitioners.  We use SIGCSE Technical Symposium (SIGCSE TS) to represent the Technical Symposium on Computer Science Education, which is commonly called SIGCSE. This differentiates the community from the conference.} added a fourth conference to its yearly schedule to address the growing need for CS education in parts of the world not served by existing conferences~\cite{Impagliazzo:2018:SLN:3287087.3287088}.  
There are many open questions in computing education that the CER community must answer~\cite{blikstein_moghadam_2019,kinnunen2010have}.

While this growth in the CER community is exciting, the real benefit comes when community members report innovations in a manner that allows other researchers to build on them and educators to appropriately apply them in their own contexts. For a community to systematically advance knowledge, its members must be able to understand published work thoroughly enough to perform replications, conduct meta-analyses, and build theories~\cite{bd12,fp04, Ahadi2016Koli,mcgill-koli2019-replication,margulieux2019measurements,Hao:2019TOCE:Replication,WWC2020,schmidt2009replication,NSF19002,gomez-reser2010-replication,ihantola2015educational}. 
In education literature, this concept is referred to as \textit{the Scholarship of Teaching and Learning} or SoTL.  
SoTL is broadly defined as ``posing [a problem] about an issue of teaching or learning, study of the problem through methods appropriate to the disciplinary epistemologies, application of results to practice, communication of results, self-reflection, and peer review''~\cite{bd12}.

In 2004, Fincher and Petre posited that most CER publications lack the type of evidence or replications required for meta-analysis or theory building~\cite{fp04}. Clear also observed that many of the standard research practices used to demonstrate rigor in  ``traditional'', non-human, computing topic areas are ignored or even actively avoided when performing CER~\cite{Clear:2006:VCS:1315803.1315806}. These observations occurred just prior to the transition to what Guzdial and du Boulay~\cite{guzdial_duboulay_2019} call the ``modern era'' of CER in 2005 with the creation of the International Computing Education Research Conference (ICER). Because this work was more than ten years ago, there is a need to understand how well more recently published research provides the type of information necessary for the computing education community to systematically advance knowledge.  Our work complements recent work in understanding the quality of reporting in CER literature~\cite{Hao:2019TOCE:Replication,ihantola2015educational, mcgill-koli2019-replication,mcgill2018improving,margulieux2019measurements,Ahadi2016Koli} and more broadly~\cite{WWC2020,NSF19002,Schulzc-BMJ2010-ConsortStatement,apa-jars} to support replication. 

Therefore, to provide insight into the current state of CER publications, the goal of this study is:

\begin{tcolorbox}[enhanced,drop shadow]
\textbf{To characterize the reporting of empiricism in Computing Education Research literature by identifying whether publications include content necessary for researchers to perform replications, meta-analyses, and theory building.}
\end{tcolorbox}

To accomplish this goal, we first defined the \textit{CER Empiricism Assessment Rubric} for analyzing CER papers.
The rubric identifies the information necessary for replication, meta-analysis, and theory building.
We then applied the rubric to 427 papers published in the \textit{Technical Symposium on Computer Science Education (SIGCSE TS)}, \textit{International Symposium on Computing Education Research (ICER)}, \textit{Conference on Innovation and Technology in Computer Science Education (ITiCSE)}, \textit{ACM Transactions on Computing Education (TOCE)} and \textit{Computer Science Education (CSE)} during 2014 and 2015 to categorize published work and identify whether information needed for replication is present and clearly labeled.
We chose these years because they were the most recent editions of the conference when we began our research. In addition, these years coincide with ten years after the first ICER~\cite{ICER14-Proceedings, ICER15-Proceedings}, a conference focused on empirical CER and just as the number of empirical research papers published at the SIGCSE TS began to increase.
Therefore, the results of this analysis will serve as a baseline against which we can compare the results of a similar analysis in subsequent years.

The contributions of this paper are:
\begin{itemize}
    \item The \textit{CER Empiricism Assessment Rubric} for evaluating the completeness of the empirical content of CER papers;
    \item An analysis of the empiricism present in CER papers published during 2014 and 2015 in five CER venues;
    \item A baseline for future analyses; and
    \item Overall observations that serve as suggestions for how the CER community can advance scientific reporting standards.
\end{itemize}

\section{Related Work}
\label{related-work}
We explore related work in three ways: 1) CER literature reviews and community reflection; 2) guidelines for reporting empirical work, particularly educational work, outside of computing; and 3) an overview of recommendations for replications.  These views of the related literature show the current efforts in transforming and increasing the impact of CER results and transfer into practice that support learners at a variety of levels and needs.  We close the section with a discussion on where our work fits into the broader literature.

\subsection{CER Literature Reviews and Community Reflection}
There have been many reviews of computing education literature, especially over the last 20 years. ~\citet{holmboe2001research} and ~\citet{Clear:2006:VCS:1315803.1315806} provide early reflections on CER and they note the biases in the broader research community, including computing, against (computing) education research as a distinct and robust research area~\cite{holmboe2001research, Clear:2006:VCS:1315803.1315806}. 
Supporting the growth and recognition of CER and  researchers through literature reviews and community reflection continues the progress made by the CER community in the last 15-20 years. 

For this section, our focus is on reviews that categorize or evaluate CER papers and reporting quality, including the use of theories and measurements. We do not include systematic literature reviews on specific topics in computing education if they do not also include some discussion on the quality of or challenges in aggregation due to reporting (e.g., reviews of K-12 ~\cite{garneli2015computing, szabo2019fifteen}; introductory programming~\cite{pears2007survey, medeiros2018systematic, vihavainen2014systematic};  teaching assistants~\cite{mirza2019undergraduate}; tools, languages, environments in K-12~\cite{mcgill2020construction}; meta-cognition and self-regulated learning~\cite{prather2020we}; and mapping of theories in CER~\cite{szabo2019review}).  
Table 1 
summarizes the reviews that include categorization or evaluation of CER papers from a quality and reporting perspective. For each review, we provide the reference information, area of focus, the years and venues considered, the number of papers reviewed, and a summary of inclusion/exclusion criteria beyond the area of focus. 
The text below also includes non-systematic literature review papers; 
however, Table 1 
does not list these papers.
We organize the discussion in each subsection in chronological order by publication date to highlight the changes in reporting since the early 2000s.

\begin{footnotesize}
\begin{longtable}{|p{.16\textwidth}|p{.20\textwidth}|p{.10\textwidth}|p{.12\textwidth}|p{.08\textwidth}|p{.18\textwidth}|}
	\caption{Summary of CER Reviews} \\
\hline
\textbf{Reference} & \textbf{Area of Focus} & \textbf{Years} & \textbf{Venues} & \textbf{\# Papers} & \textbf{Inclusion Criteria} \\
\hline
\endfirsthead

\citet{valentine2004cs}~\citeyear{valentine2004cs} & First-year CS courses including CS1 and CS2 & 1984-2003 & SIGCSE TS & 444 & CS1, CS2, and CS1/CS2 topics \\
\hline

~\citet{simon2007koli}~\citeyear{simon2007koli} & Computing education paper classification & 2001-2006 &  Koli & 102 & full papers \\
\hline
	
\citet{randolph2008methodological}~\citeyear{randolph2008methodological}  & methodological properties of reported research & 2000-2005 & SIGCSE Bulletin, CSE, JCSE, Koli, SIGCSE TS, ITiCSE, ICER, ACE & 352 & stratified random sample from all articles from listed venues between given years \\
\hline

\citet{carbone2008classifying}~\citeyear{carbone2008classifying}  & computing education paper classification & 2005-2007 & ICER & 43 & full papers \\
\hline

\citet{simon2008eight}~\citeyear{simon2008eight} & computing education paper classification & 2000-2007 & NACCQ & 157 & full papers accepted to CompEd track \\
\hline

\citet{simon2009ace}~\citeyear{simon2009ace} & computing education paper classification & 1996-2008 & ACE & 328 & full papers \\ 
\hline

\citet{joy2009categorising}~\citeyear{joy2009categorising} & categorizing types of CER literature & 2004 and/or 2005 & Too many to list & 3598 & one proceeding/volume from conference/journal \\ 
\hline

\citet{sheard2009analysis}~\citeyear{sheard2009analysis} & teaching and learning of programming & 2005-2008 & ICER, ITiCSE, SIGCSE TS, ACE, Koli, NACCQ & 164 & programming-focused paper classified as an experiment, study, or analysis \\ \hline               

\citet{kinnunen2010have}~\citeyear{kinnunen2010have} & didactic-focus-based categorization & 2005-2009 & ICER & 67 & focus on instructional process\\
\hline

\citet{malmi2010characterizing}~\citeyear{malmi2010characterizing} & CER research processes & 2005-2009 & ICER & 72 & full papers \\ 
\hline

\citet{malmi2014theoretical}~\citeyear{malmi2014theoretical} & theories, conceptual models, frameworks of CER literature & 2005-2011 & ERIC/TOCE, CSE, ICER & 308 & peer-reviewed full papers \\ 
\hline

\citet{lishinski2016methodological}~\citeyear{lishinski2016methodological}  & use of educational theory empirical results \& methodological rigor & 2012-2015 & CSE, ICER & 136 & full papers \\
\hline

\citet{ihantola2015educational}~\citeyear{ihantola2015educational} & educational data mining and learning analytics & 2005-2015 & SIGCSE TS, ICER, ITiCSE, TOCE, CSE, ACE, EDM, JEDM, Koli, PPIG, TLT & 76 & open-ended programming problems, programming process, automated data collection and analysis, length \textgreater 3 pages \\ 
\hline

\citet{al2016updated}~\citeyear{al2016updated} & empircism in CS education & 2014-2015 & SIGCSE TS & 162 & full papers with empirical results\\
\hline

\citet{luxton2018introductory}~\citeyear{luxton2018introductory}  & introductory programming   & 2003-2017 & ACM full-text collection, IEEE Explore, Science Direct (Elsevier), SpringerLink, Scopus & 1666 & introductory programming courses student in computing degrees \\ 
\hline

\citet{mcgill2018improving}~\citeyear{mcgill2018improving} & status of reporting of K-12 program elements to support  replication & 2012-2016 & SIGCSE TS, ICER, TOCE & 92  & primary and secondary education, research report or experience report, pre-college computing activities  \\ 
\hline

\citet{margulieux2019measurements}~\citeyear{margulieux2019measurements}   & standardization of measurement in CER & 2013-2017 & CSE, TOCE, ICER & 197 &       human-subjects research that measured variables excluded reviews, evaluation papers, case studies, measurement validation studies \\ 
\hline

\citet{sanders2019icer}~\citeyear{sanders2019icer} & inferential statistics & 2005-2018 & ICER & 270 & full papers \\ 
\hline

\citet{decker-sigcsets2019-instruments}~\citeyear{decker-sigcsets2019-instruments} & research-based evaluation instruments & 2012-2016 &csedreesarch.org, ICER, TOCE, CSE, Americal Evaluation Association, STELAR The Pear Institute Institute for the Integration of Technology into Teaching and Learning, MSPNet, Engineering is Elementary  & 47 & research-based evaluation instruments available in publications or public websites and databases \\
\hline

\citet{Hao:2019TOCE:Replication}~\citeyear{Hao:2019TOCE:Replication} & replications in CER literature & 2009-2018 & SIGCSE TS, ICER, ITiCSE, TOCE, CSEJTLT & 54  & articles including replicat{[}a-z{]}* manual review for replications\\ 
\hline

\citet{simon2020ace}~\citeyear{simon2020ace} & computing education paper classification & 1996-2020 & ACE & 541 & full papers \\ 
\hline

\citet{malmi2020theories}~\citeyear{malmi2020theories} & theoretical constructs and instruments for emotion, attitude, belief, and self-efficacy in computing learners & 2010-2019 & ICER, TOCE, CSE, LAK, Scopus search & 50 & computing domain containing theoretical constructs, including statistical models \\ 
\hline

\citet{papamitsiou2020computing}~\citeyear{papamitsiou2020computing} & keyword analysis of  CER literature & 2005-2019 & ITiCSE + WGR, ICER & 1274 & full papers and working group reports \\ 
\hline

\citet{sheard2020twenty}~\citeyear{sheard2020twenty} & computing education paper classification & 1996-2019 & ITiCSE + WGR & 1295 & full papers and working group  reports \\ 
\hline

This paper [2021] & quality of empirical reporting in CER & 2014-2015 & SIGCSE TS, ICER, ITiCSE, TOCE, CSE & 427 & full papers \\
\hline

\end{longtable}
\end{footnotesize}

\subsubsection{CER Topics in Literature}  
We start with an overview of how authors categorized topics in CER publications.

The seminal book by~\citet{fp04} (2004) identified 10 core areas for CER and provided an initial definition of the practices and methods of the field~\cite{fp04}. 
Their work laid the foundation for CER in the next 15 years. ~\citet{pears2005constructing} (2005) created a taxonomy of four key areas in CER that builds on and groups Fincher's and Petre's~\cite{fp04} initial categorization.  The categories included studies in teaching and learning; institutions and educational settings; problems and solutions; and CER as a discipline~\cite{pears2005constructing}.  From these categories, ~\citet{pears2005constructing} created a core CER literature including influential, seminal, and synthesis work.

\citet{joy2009categorising} (2009) created a taxonomy to categorize the types of CER publications in 21 journals and 21 conference proceedings from either 2004 or 2005.  Their findings demonstrate that conference venues tend to have a technical focus to their proceedings while journals have a pedagogical focus.

Some survey papers identified gaps in the topics covered by CER literature.  ~\citet{kinnunen2010have} (2010) reviewed 67 ICER papers between 2005 and 2009.  Their categorization considered eight categories based on a three-layered didactic structure including students, teachers, and goals/context at a classroom, organization, and societal level. Expanding their corpus of reviewed papers by 13 and considering venues beyond ICER (e.g., ACE, SIGCSE TS, PPIG, ITiCSE, ITiCSE WGR, CSE, Comput. Small Coll.), they found papers reported results at the course-level and focused on student characteristics and process.  Eight papers included additional focus areas of students' conceptions on and actions to achieve course goals and content. The literature at that time did not provide adequate coverage of categories around content/goals and teachers.

Simon and others (2007-2020) have conducted extensive classification studies of computing education literature for ACE~\cite{simon2009ace, simon2020ace}, ICER~\cite{carbone2008classifying}, ITiCSE~\cite{sheard2020twenty}, Koli~\cite{simon2007koli}, and NACCQ~\cite{simon2008eight}. The classification scheme considers the context (e.g., subject matter or course); theme (e.g., what the paper is about like teaching technique); scope (e.g., extent of collaboration and work); and nature (e.g., distinction between research and practice)~\cite{simon2007classification}.  A recent classification of ITiCSE papers in 2020 found an increase in reported research over experience reports and that teaching and learning techniques remain a common theme in published work~\cite{sheard2020twenty}. 

\citet{papamitsiou2020computing} (2020) categorized papers from 15 years of ITiCSE, ITiCSE Working Group Reports (WGR), and ICER using keywords and abstracts. The main themes identified are around introductory programming, assessments, and student performance.

The categorization of topics covered in the literature provides guidance to researchers and practitioners about areas that are well studied and ripe for meta-review and gaps where additional work is needed for a better understanding of computing education. For example, ``curriculum'' is a common categorization in many of the above surveys, however, they each have their own nuance ~\cite{fp04,joy2009categorising,kinnunen2010have,simon2007classification,papamitsiou2020computing}. While our work did not identify the context of the reviewed study (e.g., CS1, databases, etc.), we do consider the subject of the evaluation (e.g., curriculum or tool) which has overlap with categorizations from related work.  

\subsubsection{Classification by Study Type}
Several studies have classified CER literature by study type, which could include high-level categories of quantitative and qualitative or more granular categories like experiments, quasi-experimental studies, and experience reports.   

In one of the oldest CER surveys, ~\citet{valentine2004cs} (2004) examined 444 CS1/CS2 papers from SIGCSE TS between 1984 and 2003 and found that only 21\% included experimental evaluation or were experience reports.  Other categories considered were Marco Polo (``I went there and I saw this''), Philosophy, Tools, and John Henry (``outrageously difficult'') papers. There was an increase in the number of experimental papers presented in the last 10 years of the study period.

\citet{randolph2008methodological} (2008) found that of 352 papers published in various CER venues between 2000 and 2005, 40\% contained only anecdotal evidence and of the less than one-third of papers that did have experimental or quasi-experimental designs, 54.8\% of those papers used a weaker post-test only design.  ~\citet{randolph2008methodological} categorized their sampled papers against ~\citet{valentine2004cs}'s categorization and found that 40.9\% of sampled papers were experimental or experience reports.  Further, ~\citet{randolph2008methodological} used their own categorization of research methodologies and found that of the papers that reported human subject research (n = 144), 64.6\% were experimental or quasi-experimental, 26.4\% were qualitative, 18.1\% were causal comparative, 10.4\% were co-relational, and 7.6\% were survey research. 

\citet{malmi2010characterizing} (2010) identified two dimensions that described the type of research: purpose and framework.  They found that 86\% of papers had an evaluative purpose and that 79\% of papers reported a research framework.  The most common research frameworks were survey (39\%), experimental (15\%), constructive (14\%), and grounded theory (13\%).

\citet{ihantola2015educational} (2015) conducted a survey on educational data mining research published at various venues between 2005 and 2015.  Of the 76 papers that met their inclusion criteria, 78\% described studies in a natural setting.  Only 14\% reported on formal experimental research.

\citet{al2016updated} (2016) found that 162 (70\%) of the papers reviewed had some form of empirical evaluation. Their definition for empirical studies was broader than the definitions used by~\citet{valentine2004cs} and~\citet{randolph2008methodological}.
However, the detailed evaluation type numbers show that 28\% of the papers were experimental, suggesting some increase over earlier surveys.  

\citet{lishinski2016methodological} (2016) found that 71\% of CSE and 87\% of ICER papers between 2012 and 2015 reported empirical results and that 26\% from CSE and 19\% from ICER were experimental as defined by ~\citet{randolph2008methodological}.  The percentages of experimental work were much lower when considering the more specific definition of single group post-test only~\cite{lishinski2016methodological}.

Early surveys of CER literature~\cite{valentine2004cs,randolph2008methodological} reported low rates of empirical work in the surveyed papers.  More recent surveys (e.g., the last 10 years) of CER literature have shown an increase of empirical work published at a variety of CER venues. While two of the surveys reviewed literature from the CER-focused venues of ICER and CSE~\cite{malmi2010characterizing,lishinski2016methodological}, others surveys did consider other venues~\cite{al2016updated,ihantola2015educational} suggesting the increase in empiricism is happening more broadly in the computing community. However, with differences in the definitions of empirical work between surveys, direct comparisons cannot be made.

\subsubsection{Theory}
Many CER surveys discussed the use and creation of theory. 

\citet{malmi2010characterizing} (2010) found that 60\% of the surveyed papers reported explicit use of theories, models, frameworks, and instruments (TMFI) in their studies.  Additionally, they found a great diversity of TMFI used  during that time frame -- 68 distinct resources out of 78 instances.

Tenenberg's ~\cite{Tenenberg2014TOCE} (2014) position paper argues for the importance of recognizing the theories and theoretical frameworks that underlie CER research.  His argument is that authors should recognize the theoretical frameworks utilized when creating research questions because frameworks may introduce limitations on the types of questions asked and the methods used to answer research questions. Including the theoretical stance of the authors is an important aspect of reported works and serves as a foundation for future CER.

\citet{malmi2014theoretical} (2014) conducted a deeper review of CER literature that considered additional years and venues to identify the use of theories, models, and frameworks in the literature.  They found that 51\% of papers described at least one of 216 distinct theories, models, or frameworks. The interdisciplinary nature of CER benefits from the use of prior theoretical work, especially from outside fields.
However, the disparate usage of theory, models, and frameworks is a challenge for creating a ``stable theoretical base'' for CER.

\citet{lishinski2016methodological} (2016) reviewed CSE and ICER literature between 2012 and 2015 to identify the use of outside educational and learning theory and the methodological quality of the research using indicators that build on prior work (e.g.,~\cite{randolph2008methodological, malmi2014theoretical}). They identified an increase in the use of theory to support research and in the reporting of empirical CER work.
However they found many studies utilized less rigorous methods.

\citet{nelson2018use} (2018) summarize the use of theory in CER and identify three areas of concern in the community including tensions between explanation goals and design goals, lack of work on domain-specific theories, and publication bias due to theoretical lens of the work.  They also provide concrete suggestions to move the community forward by focusing on design and using theory as a guide, investing in the creation of CER-specific theories, and reducing reviewer bias in regards to manuscripts with novel designs where theory may not be appropriate.  

\citet{malmi2020theories} (2020) reviewed 50 papers with theoretical constructs around the topics of emotions, attitudes, beliefs, and self-efficacy of computing learners from a variety of venues between 2010 and 2019.  They found that three-quarters of the papers were published between 2014 and 2019 suggesting a maturation of the CER field in using theory and validated instruments. More recent papers on theoretical constructs relied on quantitative methods rather than qualitative.

There are theoretical underpinnings to CER literature and authors utilize work from outside of computing to support their research questions. More recent years have shown an increase in the use of theory to support the increased empirical work in CER.  However, the development of CER theories is an open area of future work as the field matures.  While we do not consider theory in our review of CER literature, there are enough surveys on the topic to include a discussion here for completeness.

\subsubsection{Methods \& Analysis} 
Other reviews examined the published literature to identify the type of research methods and analysis techniques used in CER studies.

\citet{randolph2008methodological} (2008) found that of the 144 studies that utilized human participants, 107 (74.3\%) used quantitative methods.  Only 15.3\% of the studies used qualitative methods and 10.4\% of studies used mixed methods.  The authors found that 44 of the 352 papers reviewed used some form of inferential statistics. Additionally, 120 of the 123 of the behavioral, quantitative, and empirical studies reported effect size.

\citet{sheard2009analysis} (2009) conducted a review of papers on the teaching and learning of programming, published in various CER venues between 2005 and 2008, which they classified as empirical.  Of the 164 papers reviewed, 37\% reported quantitative results, 21\% reported qualitative results, and 45\% utilized mixed methods.  

\citet{malmi2010characterizing} studied ICER papers published between 2005 and 2009 found a variety of analysis methods used. These methods included statistical analysis (42\%), exploratory statistical analysis (17\%), descriptive statistics (11\%), interpretive qualitative analysis (35\%), and interpretative classification or content analysis (26\%).

\citet{al2016updated} (2016) reported that 54\% the papers with empirical results in SIGCSE TS 2014 and 2015 proceedings utilized surveys and 37\% utilized experimental methods, which were typically quantitative.  Very few papers utilized qualitative methods like observations. \citet{margulieux2019measurements} (2019) found 32\% of papers published in TOCE, CSE, and ICER between 2013 and 2017 collected both qualitative and quantitative data and over half used multiple measures.  

\citet{sanders2019icer} reviewed the use of inferential statistics in ICER papers published between 2005 and 2018.  They found that 51\% of papers used inferential statistics.  However, they noted that the reporting associated with inferential statistics tended to lack precise test names, confidence levels, and p-values for results.  Other noted concerns with statistical test reporting included a lack of detail about data preparation, missing discussion about the assumptions for statistical tests, lack of explanation for ``obscure'' tests, lack of corrections for multiple statistical tests, and lack of statistical significance and effect size discussions.

The literature suggests that published CER work tends to use quantitative methods, but there is movement towards using multiple measures and appropriate statistical tests when answering research questions.

\subsubsection{Participants \& Context}
Several reviews categorized how CER literature reports on the participants of the research study and the context of the intervention(s). 

\citet{ihantola2015educational} (2015) reported that 34\% of the educational data mining studies in their survey did not report any details about course context, which could include the course level, programming language, and topics.  Additionally, 17\% of the studies did not report the number of students in the study, details about student level, or demographics.  

\citet{al2016updated} (2016) found that 11\% of papers did not report the number of participants in their study.  Additionally, when trying to identify the number of participants from the papers, there were discrepancies between reviewers due to inconsistent reporting.

\citet{mcgill2018improving} (2018) reviewed 92 papers from SIGCSE TS, ICER, and TOCE between 2012 and 2016 related to pre-college computing activities and how well papers reported key information to support future replication.  One area studied was the activity component data or information about the activities and context of the intervention.  They found that papers infrequently reported learning outcomes (25\%), curriculum (33\%), number of students in the activity (41.7\%), and details about the duration of the activity within the larger context (78.6\%), but many did not list the number of contact hours, which is important in K-12 work. They additionally found the reporting of both instructor and student demographic data lacking.  The authors provided a checklist of recommendations to support better, and more consistent, reporting of subject and context data to support replication.

\citet{luxton2018introductory} (2018) found that a ``large proportion'' of publications related to introductory programming did not provide sufficient contextual details about the reported study, which increases the difficulty for readers to determine transferability of the results.  They recommend that authors report details about the population of students and teaching context.

\citet{margulieux2019measurements} (2019) identified several empirical reporting best practices that are lacking in CER, particularly related to the context of the study. While 85\% of papers reviewed reported the number of participants only 49\% reported additional participant characteristics (e.g., prior knowledge and basic demographics).  Other contextual information that would be beneficial to include are details about the task and learning environment. 

Literature surveys that considered the context and participants in reviewed research found gaps or inconsistencies in reporting related to these items. Several reviewed papers excluded contextual information about the classroom or environment of the study~\cite{mcgill2018improving,ihantola2015educational,luxton2018introductory,margulieux2019measurements}.  Several studies found that between 10-20\% of papers did not report the number of participants~\cite{ihantola2015educational,al2016updated,margulieux2019measurements}.

\subsubsection{Data Sources}
Several studies reported on the source of the data collected, including the use of validated and non-validated instruments. 

\citet{randolph2008methodological} (2008) reported on the independent, dependent, and mediating/moderating variables in 123 behavioral, quantitative, and empirical articles in their sample of CER literature.  The most common independent variable was student instruction (98.9\%). The most common dependent variables were attitudes (60.2\%) and achievement in computer science (56.1\%). Only 29\% of papers used mediating or moderating variables, including gender, grade level, and learning styles. The identification of measures showed that questionnaires were the most commonly used (52.5\%) followed by grades (29.3\%), teacher- or researcher-made tests (22.0\%), and student work (17.9\%).

\citet{sheard2009analysis} (2009) found that empirical papers from a variety of CER venues utilized formal course assessments (42\%), tasks students complete (38\%), and questionnaires (33\%) as the top three data gathering techniques.  The authors note that the use of ``established measurement instruments'' was low in reviewed studies and that publications lacked details about how the instruments were used.

\citet{malmi2010characterizing} (2010) classified the data sources from ICER papers as (1) naturally occurring, (2) research specific data, (3) reflection, or (4) software. The results showed 79\% of those papers used research specific data, i.e. data collected specifically for the needs of the research through interviews, questionnaires, observational data, assignments, or tasks. 

\citet{al2016updated} (2016) found that most SIGCSE TS papers that used empirical evaluation reported on pedagogical techniques followed by courses and curriculum.  Additionally, they found that over 75\% of authors reported on new subjects specific to the paper. There was little replication or even reuse of data, even from the authors' previous studies.  

\citet{ihantola2015educational} (2016) found in their survey of educational data mining literature that 81\% of studies reported work from a single institution, 80\% of studies did not consider longitudinal data, and 66\% of studies only considered a single course. Additionally, they considered methods and analysis from several perspectives including details about the tasks students performed as part of the study, the type of data collected (e.g., logging, key-stroke), and the analysis methods.  They found several gaps in reporting for each of these.  

\citet{decker-sigcsets2019-instruments} (2019) consider evaluation instruments as a key data source in CER literature.  They found 47 evaluation instruments that measured cognitive, non-cognitive, and program evaluation constructs that are useful for CER researchers and would complement other quantitative and qualitative measures.  They categorized the instruments based on number of items, type of items, target demographic, reliability, and validity as reported in the literature.

The sources of data for CER work covers a variety of measures from student grades, attitudinal surveys, key-logging, and other automated data collection. A variety of evaluation instruments are available to measure various constructs~\cite{decker-sigcsets2019-instruments}. Several of the literature surveys found that data was created specifically for the research study~\cite{malmi2010characterizing,al2016updated} and focused on a single class or institution~\cite{ihantola2015educational}.  Many surveys noted that reviewed literature lacked details in reporting on the data collected and that reuse of data was low~\cite{ihantola2015educational,al2016updated}.

\subsubsection{Comparisons \& Replications} 
Several recent papers consider the impact of reporting quality on the possibility for comparisons between and replications of CER work.

\citet{ihantola2015educational} (2015) identified 5 studies (7\% of their selected papers) as replication studies in the domain of educational data mining. ~\citet{al2016updated} (2016) found that 46\% of papers reviewed had no comparison point as part of the study. The remaining papers either had a comparison to historical data or a comparison to a data set created specifically for the study. Additionally, fewer than six papers each year of the review (2014 and 2015) reported replications of prior work~\cite{al2016updated}.

\citet{Ahadi2016Koli} (2016) surveyed 73 CER researchers about their perspectives on the value of replication work in the community.  They found challenges related to the language used to describe replication/reproducibility, bias and incentives towards original work, and a lack of community value on replication work.  An example of a language challenge is related to the definitions we use for replication.  A survey quote suggests that replications are challenging due to the highly-contextualized nature of the research environment. The authors further suggest that terminology is used inconsistently in the field~\cite{Ahadi2016Koli}.  However, a replication in a new environment is a \textit{conceptual replication} and can help generalize the work~\cite{NSF19002, schmidt2009replication}.

\citet{mcgill-koli2019-replication} (2019) summarizes current work on replication, reproducibility, and meta-analysis in CER and provides a call to action on how to improve the field.  The first suggestion is to improve individual studies because replication, reproducibility, and meta-analysis rely on high quality reporting.  Other suggested actions are to pre-register studies, support open science, invest in large-scale collaborative research, report power analysis and effect size, create systems to store data and research tools, and incentivize replication through community support.  

\citet{margulieux2019measurements} (2019) found that while CER literature has adopted measurement instruments outside of computing and has created computing-specific instruments, most papers do not use standardized instruments to measure variables of interest. Use of standardized and validated instruments increases the reliability and validity of study results and will support meta-analysis by providing a common measure to compare across studies.

\citet{Hao:2019TOCE:Replication} (2019) completed a systematic literature review on replications in the CER community (including SIGCSE TS, ICER, ITiCSE, TOCE, CSEJ) between 2009 and 2018.  Of the 2,269 articles, only 54 (2.38\%) were considered a replication (as defined by~\citet{schmidt2009replication}).  Of the 54 replications, 63\% were successful replications. The others reported failures or mixed results.  Three-quarters of the replications were conceptual (e.g., methods varied from the original study), while the remaining replications were direct (e.g., methods were as similar as possible to the original study).   

CER literature does contain replications and comparisons, but replication studies are a small percentage of published CER work~\cite{Hao:2019TOCE:Replication}. 
The CER community views direct replications as challenging, with most replication studies being conceptual replications~\cite{Ahadi2016Koli,Hao:2019TOCE:Replication,schmidt2009replication,NSF19002}.  
Comparisons and replications can be supported with the use of validated instruments~\cite{margulieux2019measurements, decker-sigcsets2019-instruments} and other open science practices can support replications~\cite{mcgill-koli2019-replication}.

\subsubsection{Gaps in Reporting} 
Many literature reviews included a discussion about gaps in reporting of the literature surveyed.  These gaps are places where the literature review was more challenging because papers were missing key pieces of information.   

\citet{randolph2008methodological} (2008) reviewed 352 papers randomly selected from all papers published in several computing education venues between 2000 and 2005.  As part of their methodological review, they identified whether papers contained elements considered important by the American Psychological Association~\cite{apa-jars}, which include items like an abstract, research questions, and a description of participants.  They found that less than 50\% of the sampled papers reported purpose/rationale (36.6\%), research questions/hypotheses (22.0\%), participants description (45.5\%), procedures (37.4\%), and separate results and discussion  (29.3\%).

\citet{sheard2009analysis} (2009) report on a lack of connection between findings related to the process of learning programming and relevant theories and models of learning.  ~\citet{malmi2010characterizing} (2010) found it challenging to identify the theories, models, frameworks, and instruments utilized in the papers reviewed with similar challenges in other dimensions considered during the review. 

\citet{ihantola2015educational} demonstrated the challenges to re-analysis, replication, and reproduction resulting from a lack of context and details in existing literature through three case-studies.  Additionally, their analysis of quality measures on reviewed papers found that reporting related to confounding factors, threats to validity, and ethical issues were lacking.  Some survey responses from ~\citet{Ahadi2016Koli}'s (2016) work suggests that replications can be difficult due to gaps in the reporting of a research study.  

\citet{al2016updated} (2016) identified several areas where gaps in reporting lead to challenges when reviewing the literature particularly with regard to lack of replication and inconsistent paper organization. In particular, they found that between 40\% of 70\% of papers in the years evaluated from the SIGCSE TS were lacking threats to validity.

\citet{lishinski2016methodological} (2016) found that 47\% of CSE and 56\% of ICER papers during their study timeframe reported explicit research questions.  While this level of reporting is an improvement over the use of research questions as reported by ~\citet{randolph2008methodological}, the lack of explicit research questions in some empirical studies is a concern.  ~\citet{lishinski2016methodological}'s finding about research questions are lower than the 79\% reported by ~\citet{ihantola2015educational} possibly due to the focus on educational data mining and learning analytics.

\citet{mcgill2018improving} (2018) found many gaps in reporting of pre-college activities related to the study context and the demographics of instructors and students.  They provide a checklist with recommendations that would also be useful for education researcher more broadly.

\citet{luxton2018introductory} (2018) found in their review of introductory programming that many papers lacked details about the construct/intervention and its operationalization and effect sizes, which limits replication. They suggest archiving course information including ``syllabus, learning outcomes, infrastructure, teaching approaches, and population demographics...''.

Prior work found several gaps in reporting quality~\cite{randolph2008methodological,ihantola2015educational,al2016updated,lishinski2016methodological}, lack of connection between results and theories~\cite{sheard2009analysis,malmi2010characterizing}, and lack of contextual information~\cite{ihantola2015educational,mcgill2018improving,luxton2018introductory}.
Open science practices, like archiving research study information, could supplement the details in the literature~\cite{luxton2018introductory,mcgill-koli2019-replication}.

\subsection{Guidelines for Reporting and Quality} \label{reporting-quality}
Several organizations provide guidelines for reporting and assessing study quality. 
We encourage interested readers to review these resources to consider additional quality expectations when designing and reporting on empirical CER studies.

The \textit{What Works Clearinghouse} (WWC) has developed a \textit{Standards Handbook}~\cite{WWC2020} that provides resources for a systematic review process, including the development of a quality rubric, to provide summary reports about empirical educational research that can be of use to policymakers. The structured review is intended to assess the \textit{internal validity} of the study.  WWC focuses on four types of research: randomized controlled trials, quasi-experimental design, regression discontinuity design, and single-case design.  Studies related to a particular area of interest and quality rubric provide one of three labels: meets WWC Design Standards without reservations, meets WWC Design Standards with reservations, does not meet WWC Design Standards.  For example, quasi-experimental designs which lack randomization can only achieve the category ``meets WWC Design Standards with reservations'' due to the lack of randomization in the process. The WWC Design Standards provide a deeper quality assessment, specifically around the methods, results, and internal validity of educational research studies than our rubric on reporting norms.  

The CONSORT 2010 Statement~\cite{Schulzc-BMJ2010-ConsortStatement} provides guidelines and a checklist for reporting parallel group randomized trials. 
While the authors of CONSORT do not prescribe a specific article structure, they do suggest using subheadings to help readers find information in the manuscript.  
Appropriate subheadings provide readers with guideposts on where to find important information and can be especially useful for supporting possible replications.

The American Psychological Association (APA) provides \textit{Journal Article Reporting Standards (JARS)} for researchers, reviewers, and editors~\cite{apa-jars}. The website provides guidelines for reporting standards on quantitative, qualitative, and mixed research publications and study designs. It also provides guidance on meta-analysis reporting standards.  

The American Education Research Association (AERA) \textit{Standards for Reporting on Empirical Social Science Research in AERA Publications} provides guidance for reporting on empirical educational research~\cite{area-empirical-social-science-2006}.  Guidance includes a clear statement of the problem, description of the study design, overview of the data collection and sources of evidence, a description of key measurements and how they are operationalized by the study, details of analysis procedures, scope, and ethics.

Within the computing education community, several articles provide guidance on high quality reporting of CER work. ~\citet{daniels2012models} provide a framework for designing action research to support answering research questions about ``concrete teaching and learning challenges'' in computer science classrooms.  Their framework provides guidance on reporting the researcher's epistemology and theoretical perspective(s), describing the context of the research study, and describing the methodology and methods associated with the research question. ~\citet{mcgill-fie2018-cer-repository} describe the challenges in creating a repository for K-12 CER research due to inconsistencies in reporting standards and questions about study quality identified by a focus group gathered to create repository requirements. Some of the findings from the focus group suggest that reporting on basic technical components would support replication.  
The \url{www.csedresearch.org} website serves as a repository for instruments and guidelines for reporting on K-12 studies, which builds on other work by McGill, Decker, and others~\cite{mcgill2018improving}. As part of their systematic literature review, ~\citet{ihantola2015educational} created a quality assessment rubric for selected studies that reported on educational data mining and learning analytics.

Researchers in related fields which also consider human subjects, like software engineering, have proposed reporting guidelines for empirical work. ~\citet{runeson2009guidelines} describes guidelines for reporting case studies, including suggested headings and subsections.  The work~\cite{runeson2009guidelines} compares case study reporting guidelines to experimental reporting guidelines created by ~\citet{jedlitschka2005reporting} and refined by ~\citet{kitchenham2008evaluating}.  
Additionally, ~\citet{kitchenham2008evaluating} provides checklists that researchers, practitioners, meta-reviewers, replicators, and reviewers should consider when reading experimental work. ~\citet{Carver-RESER2010-ReportingGuidelines} provides guidelines for reporting experimental replications in software engineering that can also be useful as general reporting guidelines for empirical work.  As part of reporting on a replication, the original study should be discussed, including the research question(s), participants and their characteristics and context, experimental design, artifacts or resources used in the study, context variables, and summary of the results as general reporting guidelines for empirical work. 

\subsection{Guidelines for Replication}
Our focus on reporting quality is in the goal of supporting replications, meta-analysis, and theory-building in the CER community.  This section describes key references on replications in CER and more broadly.

\citet{schmidt2009replication} provides a pragmatic and operational overview of replication, particularly in the social sciences.  Their definitions of direct replication and conceptual replication form the foundation of the discussion of replication and capture our views of replication. Replications provide additional confirmatory power about a research question.  Schmidt suggests five functions of replication: ``to control for sampling error, to control for artifacts, to control for fraud, to generalize results to a larger or to a different population, and to verify the underlying hypothesis of the earlier experiment.''  While \citet{schmidt2009replication} provides a framework for replications and reproductions, he additionally provides a pragmatic reflection on practical publications of reflections given community biases towards original work by using \textit{follow-up studies} and \textit{systematic replications}.  Follow-up studies may include direct replication of earlier work and additional experimental conditions that either provide generalizability or novelty beyond the replicative piece.  Systematic replications provide a way of exploring the variation around a research replication modeled as a data matrix.  Replications consider different cases within the matrix to systematically explore the research space.

The National Science Foundation (NSF) and the Institute of Education Sciences (IES) issued joint \textit{Companion Guidelines on Replication \& Reproducibility in Education Research}~\cite{NSF19002} as a supplement to the \textit{Common Guidelines for Education Research and Development}~\cite{NSF13126}. The \textit{Companion Guidelines} define reproducibility and replication and describe the importance of both in furthering theory in education.  The guidelines for designing reproducible studies suggest that ``Analyses should be described in sufficient detail as to allow other researchers to reproduce the results using the same dataset.''  When reporting results, researchers should share data and analysis details, how results compare to replicated or reproduced studies, and specific details about data that was excluded or omitted~\cite{NSF19002}.  The definitions from the \textit{Companion Guidelines}, which are similar to those provided by ~\citet{schmidt2009replication}, are:

\begin{itemize}
    \item \textit{reproducibility}: ``the ability to achieve the same findings as another investigator using extant data from a prior study''~\cite{NSF19002}.
    \item \textit{replication}: ``involved collecting and analyzing data to determine if the new studies (in whole or in part) yield the same findings as a previous study''~\cite{NSF19002}. Replication studies are further broken down into two categories:
    \begin{itemize}
        \item \textit{direct replication}: ``seek to replicate findings from a previous study using the same, or as similar as possible, research methods and procedures as a previous study''~\cite{NSF19002,schmidt2009replication}.
        \item \textit{conceptual replication}: ``seek to determine whether similar results are found when certain aspects of a previous study's method and/or procedures are systematically varied''~\cite{NSF19002,schmidt2009replication}.
    \end{itemize}
\end{itemize}

\citet{nosek2014method} describe the benefits of direct replication as increasing the 
size of the data that can help identify false positive results, help establish generalizability of results, and identify the boundaries of results. 

\citet{gomez-reser2010-replication} explore the language and definitions of how other disciplines verify findings of experimental research.  A review of literature found 18 sources with classifications of replications.  Their synthesis identified three key groupings of definitions based on whether the authors use the \textbf{same methods} for either \textit{operational replication} of the methods or \textit{empirical generalization} of the results, \textbf{different methods} for a \textit{conceptual replication} to determine if the results around a given hypothesis are reproducible, or \textbf{existing data sets} where the analysis is replicated for \textit{internal replication} or for a \textit{reanalysis of the data} using different methods. ~\citet{gomez-reser2010-replication} provide the following terms and definitions:

\begin{itemize}
\item \textit{Re-analysis}: uses the same or different analysis on ``the data of a previously run experiment are used to verify the results rather than re-running the experiment.'' ~\cite{gomez-reser2010-replication}
\item \textit{Replication}: uses the same methods with different data that ``verifies that the observed findings are stable enough to be discovered more than once.'' ~\cite{gomez-reser2010-replication}
\item \textit{Reproduction}: uses the same hypothesis, but different methods and data to verify ``that the findings are not to be attributed to the experimental method.''~\cite{gomez-reser2010-replication}
\end{itemize}

\citet{ihantola2015educational} build on G{\'o}mez et al.'s ~\cite{gomez-reser2010-replication} work to create a novel classification of reproduction studies in the R.A.P. Taxonomy.  They consider three key criteria: \textit{researchers}, \textit{data analysis}, and \textit{data production} to identify seven classifications where one or more of the criteria are changed for study reproduction.  By considering the taxonomy, researchers can examine the state space of possible study reproductions \textit{related to a specific hypothesis}. The seven categories and their definitions are:

\begin{itemize}
\item \textit{Re-Analysis}: ``a different different experimenter is following the same analysis done before with the original data set for review purposes''~\cite{ihantola2015educational}
\item \textit{Extended Analysis}: ``an experimenter is extending the baseline study by looking at a previously analyzed data set, but using new analysis methods~\cite{ihantola2015educational}
\item \textit{Repetition}: ``an experimenter is repeating the same analysis with new data''~\cite{ihantola2015educational}
\item \textit{Verification}: ``the same data set is looked into again by a different experimenter and using a different analysis method to verify the conclusions''~\cite{ihantola2015educational}
\item \textit{Replication Study}: ``a different experimenter is following the same analysis method as in the baseline study, but using their own different data set''~\cite{ihantola2015educational}
\item \textit{Triangulation}: ``an experimenter is collecting a new data set to be analyzed with a new method''~\cite{ihantola2015educational}
\item \textit{Reproduction}: ``a different experimenter is analyzing their own new data set and following a new analysis method designed for the study in order to test the hypothesis in the baseline study.''~\cite{ihantola2015educational}
\end{itemize} 

ACM updated their Artifact Review and Badging definitions~\cite{acm-artifact-review} in 2020 to correspond to definitions used by the National Information Standards Organization (NISO). ACM uses the following definitions:

\begin{itemize}
\item \textit{Repeatability}: ``Same team, same experimental setup''~\cite{acm-artifact-review}
\item \textit{Reproducibility}: ``Different team, same experimental setup''~\cite{acm-artifact-review}
\item \textit{Replicability}: ``Different team, different experimental setup''~\cite{acm-artifact-review}
\end{itemize}

The terminology utilized by NSF and ISE \textit{Companion Guidelines}~\cite{NSF19002}, ~\citet{gomez-reser2010-replication}, ~\citet{ihantola2015educational}, and ACM~\cite{acm-artifact-review} is inconsistent . Table~\ref{table-definitions} maps the definitions to each other. We utilize the definitions from the \textit{Companion Guidelines}~\cite{NSF19002} in our discussion.

\begin{table}[!htb]
\centering
\caption{Alignment of Replication Definitions}
\label{table-definitions}

\begin{tabular}{|l|l|l|l|}
\hline
\textbf{\begin{tabular}[c]{@{}l@{}}NSF/ISE \\ Companion Guidelines\end{tabular}~\cite{NSF19002}} & \textbf{G{\'o}mez, et al.}~\cite{gomez-reser2010-replication} & \textbf{Ihantola, et al.}~\cite{ihantola2015educational}  &  \textbf{ACM}~\cite{acm-artifact-review}                 \\ \hline
Reproducibility                       & Re-analysis                                                                 & \begin{tabular}[c]{@{}l@{}}Re-analysis\\ Verification\end{tabular}  &  Repeatability                    \\ \hline
Direct Replication                    & Replication                                                                 & \begin{tabular}[c]{@{}l@{}}Repetition\\ Replication Study\end{tabular}      & Reproducibility             \\ \hline
Conceptual Replication                & Reproduction                                                                & \begin{tabular}[c]{@{}l@{}}Extended Analysis\\ Triangulation\\ Reproduction\end{tabular} & Replicability \\ \hline
\end{tabular}

\end{table}

\subsection{Contributions}
The recent increase in characterizing the reporting standards in CER literature through various reporting lenses demonstrates the need for the community to identify and use common reporting standards. 
We see our contribution as complementing the recent work of others~\cite{Hao:2019TOCE:Replication,sanders2019icer,margulieux2019measurements,ihantola2015educational,mcgill2018improving,lishinski2016methodological,Ahadi2016Koli,al2016updated,mcgill-koli2019-replication,mcgill2020construction,malmi2020theories,nelson2018use,Tenenberg2014TOCE} to improve reporting standards in the CER community.
Specifically, the current paper provides an advance over prior work in three dimensions:
\begin{itemize}
    \item \textit{Scale} - we considered the proceedings of five CER venues for a full two years, including the SIGCSE TS;
    \item \textit{Scope} - our study provides a broad snapshot that covers the full proceedings/issues of CER venues compared with the more focused snapshots of prior work.
    \item \textit{Focus} - our study investigates the quality of empirical reporting in CER rather than classification.
\end{itemize}

\section{Methodology}
\label{methodology}
Our goal is to describe the reporting of empiricism across the most common venues for computing education research and practice. 
We have identified elements of empiricism that should be included in reports of research results, which parallel the key items needed during design of an educational research study~\cite{fp04,bd12,creswell_creswell_2018}.
Our systematic literature review~\cite{kitchenham2004slr} will reveal the  state of the practice for reporting empirical research results and provide guidance to the CER community for how to improve reporting to provide the basis the community needs to advance in a scientifically rigorous manner.

Our study investigates the following research questions (RQs):

\begin{itemize}
    \item RQ1: What percentage of papers in CER venues have some form of empirical evaluation? 
    \item RQ2: Of the papers that have empirical evaluation, what are the characteristics of the empirical evaluation? 
    \item RQ3: Of the papers that have empirical evaluation, do they follow norms (both for inclusion and for labeling of information needed for replication, meta-analysis, and, eventually, theory-building) for reporting empirical work? 
\end{itemize}

\subsection{Selection Criteria}
Previous work done through 2008 \cite{randolph2008methodological, valentine2004cs} showed a small increase in the presence of empirical evaluation in papers in CER conferences in general.
Several more recent papers show an increase in empirical work across a variety of venues~\cite{malmi2010characterizing,al2016updated,lishinski2016methodological,ihantola2015educational}.
Our evaluation includes two years of papers from five venues (SIGCSE TS, ICER, ITiCSE, TOCE, and CSE).
We included all papers from the TOCE and CSE editions and two iterations of each conference during 2014 and 2015.

\subsection{CER Empiricism Assessment Rubric}
We developed and piloted the first version of the \textit{CER Empiricism Assessment Rubric} in 2015~\cite{al2016updated}.  
This initial version of the rubric was based on the evaluation rubrics used in previous reviews of CER literature~\cite{randolph2008methodological, valentine2004cs} and on items considered when designing empirical studies~\cite{fp04,bd12}.
Specifically, the rubric was focused on aspects of a research paper that are essential for others to understand and potentially replicate the experiment, as evidenced by prior work~\cite{randolph2008methodological, valentine2004cs} and supported by guidelines in educational research~\cite{WWC2020, Schulzc-BMJ2010-ConsortStatement, NSF19002, schmidt2009replication}.
Randolph et al.'s rubric, however, incorporated more granular details in its categories.  
For example, Randolph et al.'s rubric took into account the computing domain from which the research intervention originated, including categories such as ``Visualization'' and ``Simulation'' as discrete options~\cite{randolph2008methodological}.
Our rubric takes a more high-level approach, focusing more on the general role of the intervention in how it relates to computing education.
In this example, instead of noting whether the intervention incorporated ``Visualization'' or ``Simulation,'' we focused on whether the intervention was an assignment, a tool, a curricular innovation, etc.  

After we applied the initial version of the rubric in our study of the SIGCSE TS proceedings, we evolved both the rubric and the methodology for applying it to a large number of papers to arrive at the current version of the rubric found in Appendix A.
In updating our rubric, we referenced the evolving reviewing guidelines for ACM SIGCSE conferences and work in the scholarship of teaching and learning (SoTL) space to ensure that our core rubric items were capturing current best practices in CER work~\cite{bd12}.
Further, we streamlined our evaluation methodology to make the application of the rubric to a large set of papers more feasible, while still maintaining the quality of our evaluation by examining the inter-rater reliability.
Finally, we discovered some papers that exposed corner cases that we needed to address, such as papers that purported to be empirical studies, but reported no number of actual participants in the study.

The rubric is comprised of a \textit{Base Rubric} that is applicable to all
papers with some level of empirical results and three additional rubrics for specific categories of research projects.

The \textit{Base Rubric} contains a set of characteristics that captures the overall nature of the empirical work including information such as the type of evaluation method, the evaluation subject, whether there is any comparison, and the number of participants.  
These items provide a high-level characterization of the empirical work presented, answering questions such as:
\begin{itemize}
    \item What is the balance between quantitative and qualitative work?
    \item At what rate do researchers publish studies on curricula?
    \item Are researchers performing replication studies or creating their own interventions?
    \item How many participants are in the typical CER study?
\end{itemize}

Depending upon the type of empirical study included in the paper, i.e. quantitative, qualitative, survey, and/or descriptive, we then employed one or more additional rubrics.
Each of these rubrics allowed us to characterize the presence and clarity of important information that CER papers should include relative to each type of study.
The items on these additional rubrics all use the same categorization scale that identifies whether a piece of information is \textit{present} and how easy it is for the reader to \textit{identify} that piece of information.
The scale captures both types of information as follows:
\begin{itemize}
    \item{\textit{Completeness -  the level of completeness of presented information}}
    \begin{itemize}
        \item Complete - Answers/addresses all questions for a rubric category.  There is no assessment on the quality of the answer.
        \item Partial - Answers/addresses some of the questions for a rubric category.  There is no assessment on the quality of the answer.
        \item Not Present - Answers/addresses none of the questions for a rubric category. The questions should  be addressed.
        \item Not Applicable - The rubric item is not applicable to the paper.
    \end{itemize}
    \item{\textit{Labeled - whether the presented information is clearly labeled}}
        \begin{itemize}
        \item Labeled - There is a heading appropriate for the rubric item or there is emphasis (bold/italics) for the rubric item.
        \item Not Labeled - There is no heading or emphasis for the rubric item to easily find the item in the paper.
    \end{itemize}
\end{itemize}

Therefore each rubric item can take one of six labels:
\begin{itemize}
    \item Complete and Labeled
    \item Complete and Not Labeled
    \item Partial and Labeled
    \item Partial and Not Labeled
    \item Not Present
    \item Not Applicable
\end{itemize}

The items included on the rubric represents the type of information necessary to support reproducibility and replication of research studies. We use the definitions from the \textit{Companion Guidelines on Replication \& Reproducibility in Education Research} published jointly by the National Science Foundation (NSF) and the Institute of Education Sciences (IES)~\cite{NSF19002}.

\begin{itemize}
    \item \textit{reproducibility}: ``the ability to achieve the same findings as another investigator using extant data from a prior study''~\cite{NSF19002}.
    \item \textit{replication}: ``involved collecting and analyzing data to determine if the new studies (in whole or in part) yield the same findings as a previous study''~\cite{NSF19002}. Replication studies are further broken down into two categories:
    \begin{itemize}
        \item \textit{direct replication}: ``seek to replicate findings from a previous study using the same, or as similar as possible, research methods and procedures as a previous study''~\cite{NSF19002,schmidt2009replication}.
        \item \textit{conceptual replication}: ``seek to determine whether similar results are found when certain aspects of a previous study's method and/or procedures are systematically varied''~\cite{NSF19002,schmidt2009replication}.
    \end{itemize}
    \item \textit{meta-analysis} - is a quantitative, formal, systematic analysis of a number of related study results in order to identify general patterns and draw overarching conclusions about the entire body of research~\cite{Haidich:2010}. 
    \item \textit{theory-building} - is the process of using replication, repetition, and meta-analysis to systematize knowledge. Theories can drive the generation of hypotheses and produce predictive theory through scientific enquiry~\cite{fp04}.  Additionally, conceptual replications can produce understanding and confirm underlying theory~\cite{schmidt2009replication}.
\end{itemize}

For the remainder of this paper, we use the term \textit{replication} to cover any type of study design that builds on prior study designs through replication in the same setting (i.e., direct replication) or through systematic variance (i.e., conceptual replication).

Reporting common information and using open science supports replications~\cite{NSF19002}.  
Meta-analysis can then utilize these common results, aggregate them, and increase overall generalizability of findings to ``gain a better understanding of what interventions improve (or do not improve) educational outcomes, for whom, and under what conditions''~\cite{NSF19002}. Because individual studies are more prone to error or bias, meta-analysis or meta-studies provide the opportunity to synthesize results around a research question to understand impacts on a broader set of learners~\cite{mcgill-koli2019-replication,al2016updated,schmidt2009replication,NSF19002}.  
This process provides for a mechanism for theory, either confirmatory of educational theory in a computing context or the emergence of theory to support new phenomenon~\cite{fp04,schmidt2009replication,mcgill-koli2019-replication}.

For example, good practice is for a paper to provide an overview of the participants, including demographics and the sampling or recruitment method~\cite{mcgill2018improving}.
If a paper met all the criteria specified in the rubric, the paper receives a label of ``Complete'' for that rubric item.  
If the paper meets at least one, but not all, of the criteria, the paper receives a label of ``Partial'' for that rubric item. 
If the paper meets none of the criteria in the rubric, the paper receives a label of ``Not Present'' for the rubric item.  
Finally, if this particular rubric item does not apply to the paper, the paper receives the label of ``Not Applicable'' for that rubric item.

To ease the replication, meta-analysis, or theory-building process, it is important for a paper to not only contain the appropriate information but also to present that information in a way that readers can easily locate it, similar to the checklists in CONSORT~\cite{Schulzc-BMJ2010-ConsortStatement} and suggested information for reporting in APA JARS~\cite{apa-jars}.
We determined a rubric item to be easy to find if there was a relevant section heading in the paper for the item.  
We also considered text that emphasized a specific item through italics, bolding, or a bulleted list. These are common ways of noting research questions rather than using a section header.
We assigned a value of either ``Labeled'' or ``Not Labeled'' for each rubric item in each paper.

To reiterate, the purpose of the \textit{CER Empiricism Assessment Rubric} is to characterize the study and the rigor in which a paper reports an empirical study.
The goal is to comment on the efficacy of the reporting from the perspective of a reader who wishes to replicate, perform meta-analysis, or build theories.
The rubric does not comment in any way on the quality of the study or reporting itself, beyond the presence of the expected information.

We completed the \textit{Experimental Rubric} for each paper that contained some form of empirical evaluation.
This rubric characterizes the rigor in which the paper reports and labels the key aspects of an empirical study~\cite{Schulzc-BMJ2010-ConsortStatement,fp04,bd12,creswell_creswell_2018}.
These aspects include concepts and questions such as:
\begin{itemize}
    \item Are the research objectives obvious and easy to find?
    \item Do the authors present related work?
    \item Is the study design properly presented?
    \item How was the data gathered and analyzed for this work?
    \item Are the results presented in a succinct, direct way?
    \item Are threats to validity discussed?
\end{itemize}
In order for an empirical study to be replicated, a paper needs to thoroughly discuss and clearly label these items so readers can easily find them.

We completed the \textit{Survey Rubric} for each paper that uses a survey as either the primary research methodology or one of the data sources.  
Within those topics, 
the rubric items examine whether the paper describes the survey creation process, the rationale behind the questions, and the execution of the survey, including its administration and the medium used.

We completed the \textit{Descriptive/Persuasive Rubric} for papers that do not claim any cause/effect relationship.
These papers are often presented as position papers.
This rubric contains items focused on the goal of the paper's argument, the presence of related work, the soundness of the argument, and whether supporting evidence is present.

\subsection{Gathering Data and Applying the Rubric}
Because we included all papers from each venue in the chosen years, we manually extracted the DOI entries for each paper from the digital library.
Then we downloaded the PDFs of the papers to analyze locally.

In an initial phase, we examined the first 50 papers to refine our methodology.
We randomly assigned each paper to two researchers for initial categorization.
Each researcher independently evaluated the paper using the rubric.
Then we met to discuss any discrepancies in the analysis.
If needed, we asked a third researcher to review a paper to resolve any discrepancies. 
At the end of this process, all researchers agreed on the final categorization of the paper.

Based on the experience with the first 50 papers, we modified our methodology for the remaining papers.  
Overall, we found very few discrepancies in this first set of papers. 
We computed kappa to measure the inter-rater reliability for each of the rubric items across the 50 papers.
Overall, our level of agreement was very high, with the kappa values for each rubric item above 0.8 (p \textless 0.01).
Based on this high level of agreement, we decided it was not necessary for two reviewers to complete the full rubric on each paper.
However, we also did not want to fully rely on only one reviewer for each paper.
As a compromise, because the \textit{Base Rubric} contains the information that determines the type of research present in the paper and determines which other rubric(s) apply, we had two researchers complete the \textit{Base Rubric} for each paper.
After resolving any discrepancies, this process resulted in a consensus on the type of research present in each paper.
Based on the type of research present, one researcher then reviewed the paper using one or more of the remaining rubrics (\textit{Experimental, Survey, and Descriptive/Persuasive Rubrics}).
Finally, we merged all results into a single spreadsheet for analysis and calculations~\cite{dataset_2021}.

\section{Results}
\label{results}
This section describes the results of our analysis of our dataset~\cite{dataset_2021}, organized around the three research questions.

\subsection{RQ1: Empirical Evaluation}

\begin{tcolorbox}[enhanced,drop shadow]
\textbf{RQ1: What percentage of papers in CER venues have some form of empirical evaluation?}
\end{tcolorbox}

We classify a paper as \textit{empirical} if its \textbf{Evaluation Method} is \textit{Descriptive}, \textit{Survey}, \textit{Qualitative}, or \textit{Quantitative} such that it contains \textit{empirical evidence} addressing a research goal, question, or hypothesis.
As shown in Table~\ref{tab_empirical_evaluation}, 351 of the 427 papers (82\%) contained some type of empirical evaluation. 
While this number is quite high, note that we did not assess the quality or correctness of the empirical evaluation, just the presence of it.
When considering specific venues, we found that TOCE had a lower percentage of empirical papers.
This result is likely due to the fact that a portion of the TOCE papers were editorials and opinion papers, which we would not expect to contain empirical evaluation.
On the other end of the spectrum, 24 out of 25 papers in ICER 2015 had some sort of empirical evaluation.

\begin{table}[!htb]
\centering
\caption{Empirical Evaluation by Venue and Year}
\label{tab_empirical_evaluation}
\begin{tabular}{|l|r||r|r||r|r||r|}
\hline
\textbf{Venue}  & \multicolumn{1}{l||}{\textbf{Year}} & \multicolumn{1}{l|}{\textbf{\# Papers}} & \multicolumn{1}{l||}{\textbf{\begin{tabular}[c]{@{}l@{}}Acceptance \\ Rate\end{tabular}}} & \multicolumn{1}{l|}{\textbf{\begin{tabular}[c]{@{}l@{}}\# Empirical \\ Papers\end{tabular}}} & \multicolumn{1}{l||}{\textbf{\begin{tabular}[c]{@{}l@{}}\% Empirical\\   Papers\end{tabular}}} & \multicolumn{1}{l|}{\textbf{\begin{tabular}[c]{@{}l@{}}\# Empirical \\ Papers \\ by Venue\end{tabular}}} \\ \hline
SIGCSE TS       & 2014                               & 110                                     & 39.4\%                                                                                   & 91                                                                                           & 82.7\%                                                                                        & \multirow{2}{*}{176}                                                                                     \\ \cline{1-6}
SIGCSE TS       & 2015                               & 103                                     & 36.0\%                                                                                   & 85                                                                                           & 82.5\%                                                                                        &                                                                                                          \\ \hline
ICER            & 2014                               & 17                                      & 25.0\%                                                                                   & 14                                                                                           & 82.4\%                                                                                        & \multirow{2}{*}{38}                                                                                      \\ \cline{1-6}
ICER            & 2015                               & 25                                      & 26.0\%                                                                                   & 24                                                                                           & 96.0\%                                                                                        &                                                                                                          \\ \hline
ITiCSE          & 2014                               & 51                                      & 35.0\%                                                                                   & 43                                                                                           & 84.3\%                                                                                        & \multirow{2}{*}{85}                                                                                      \\ \cline{1-6}
ITiCSE          & 2015                               & 52                                      & 43.5\%                                                                                   & 42                                                                                           & 80.8\%                                                                                        &                                                                                                          \\ \hline
TOCE            & 2014                               & 21                                      & n/a                                                                                      & 15                                                                                           & 71.4\%                                                                                        & \multirow{2}{*}{29}                                                                                      \\ \cline{1-6}
TOCE            & 2015                               & 21                                      & n/a                                                                                      & 14                                                                                           & 66.7\%                                                                                        &                                                                                                          \\\hline
CSE            & 2014                               & 11                                      & n/a                                                                                      & 10                                                                                           & 90.9\%                                                                                        & \multirow{2}{*}{23}                                                                                      \\ \cline{1-6}
CSE            & 2015                               & 16                                      & n/a                                                                                      & 13                                                                                           & 81.3\%                                                                                        &                                                                                                          \\ \hline \hline
\textbf{Totals} & \multicolumn{1}{l||}{}              & \textbf{427}                            & \textbf{}                                                                                & \textbf{351}                                                                                 & \textbf{82.2\%}                                                                               & \multicolumn{1}{l|}{}                                                                                    \\ \hline
\end{tabular}
\end{table}

When comparing these results with those from earlier literature reviews, we see an increase in the percentage of papers containing empirical evaluation over time. ~\citet{valentine2004cs} found 21\% of papers discussing CS1 and CS2 topics had ``experimental'' evaluation and ~\citet{randolph2008methodological} found  35\% of papers with behavioral, quantitative, or empirical research from a broader set of CS education venues from 2000-2005 contained empirical evaluation.  Early surveys of ICER papers between 2005-2009 found 86\% of papers had an evaluative purpose, appropriate for a conference focused on CER~\cite{malmi2010characterizing}. ~\citet{lishinski2016methodological}'s analysis considered ICER and CSE papers between 2012 and 2015 and found that 71\% of CSE papers and 87\% of ICER papers reported empirical results. These numbers are lower than some of our results but consider a larger time frame. A review of SIGCSE TS papers from 2014 and 2015 found that 70\% of papers had some form of empirical evaluation~\cite{al2016updated}. Our definition of empirical evaluation is more broad resulting in higher numbers during the same time period. When looking at data mining and automated learning research across venues, ~\citet{ihantola2015educational} found that 78\% of papers reported on a study in a natural (e.g., empirical) setting.
Our findings, along with complementary work, suggest an increase in the amount of empirical work since the early surveys by ~\citet{valentine2004cs} and ~\citet{randolph2008methodological}, which in turn should lead to an increase of CER in the SIGCSE community.
The remaining research questions investigate the state-of-the-practice for reporting empirical studies to further understand the concerns raised in our previous work~\cite{al2016updated}.

\subsection{RQ2: Characteristics of Empirical Evaluation}
\begin{tcolorbox}[enhanced,drop shadow]
    \textbf{RQ2: Of the papers that have empirical evaluation, what are the characteristics of empirical evaluation?}
\end{tcolorbox}

To answer this question, we characterize the papers using the \textit{CER Empiricism Assessment Rubric} -- \textit{Base Rubric}. 
For each characteristic, we provide a table that reports the raw number of papers with that characteristic type followed by the percentage of the total empirical papers for the venue (based on the last column of Table \ref{tab_empirical_evaluation}).  
We gave some papers multiple values for a characteristic.
In those cases, the percentages may total more than 100\%. Additionally, we summarize the characteristic items in the total column as the percentage of all 351 empirical papers.

\begin{table}[!htb]
\centering
\caption{Evaluation Methods by Venue}
\label{tab_evaluation_method}
\begin{tabular}{|l||r|r||r|r||r|r||r|r||r|r||r|r|}
\hline
\textbf{\begin{tabular}[c]{@{}l@{}}Evaluation\\ Method\end{tabular}}                 & \multicolumn{2}{c||}{\textbf{\begin{tabular}[c]{@{}c@{}}SIGCSE \\ TS\end{tabular}}} & \multicolumn{2}{c||}{\textbf{ICER}} & \multicolumn{2}{c||}{\textbf{ITiCSE}} & \multicolumn{2}{c||}{\textbf{TOCE}} &
\multicolumn{2}{c||}{\textbf{CSE}} &
\multicolumn{2}{c|}{\textbf{TOTAL}} \\ \hline 
\textbf{}                       & \textbf{\#}       & \textbf{\%}      & \textbf{\#}      & \textbf{\%}     & \textbf{\#}       & \textbf{\%}      & \textbf{\#}      & \textbf{\%}  &\textbf{\#}      & \textbf{\%}  &\textbf{\#} & \textbf{\%}   \\ \hline \hline
\textbf{\begin{tabular}[c]{@{}l@{}}Descriptive/\\ Persuasive\end{tabular}} & 7                 & 4\%            & 0                & 0\%           & 1                 & 1\%            & 2                & 7\% & 3                & 13\%       & 13 & 4\%    \\ \hline
\textbf{Survey}                 & 32                & 18\%           & 4                & 11\%          & 12                & 14\%           & 7                & 24\% & 4                & 17\%     & 59 & 17\%     \\ \hline
\textbf{Qualitative}            & 15                & 9\%            & 12               & 32\%          & 13                & 15\%           & 4                & 14\% & 5                & 22\%    & 49 & 14\%     \\ \hline
\textbf{Quantitative}           & 126               & 72\%           & 22               & 58\%          & 59                & 69\%           & 17               & 59\% & 12                & 52\%     & 236 & 67\%    \\ \hline
\end{tabular}
\end{table}

\hfill \break
\noindent
\textbf{Evaluation Method (Table~\ref{tab_evaluation_method}).}
The \textit{CER Empiricism Assessment Rubric} describes each evaluation method in detail. 
However, for clarity we highlight the \textit{Survey} and \textit{Descriptive/Persuasive} evaluation methods more carefully here to describe their specific usage as an evaluation method. 
We coded papers as \textit{Survey} when they described a community survey with the goal of describing the current state of that community (i.e. the paper does not involve any interventions).  
\textit{Qualitative} and \textit{Quantitative} evaluation methods may have surveys as one of their \textit{Data Sources} in the study.   
We coded papers as \textit{Descriptive/Persuasive} if its focus was to describe a current situation or to persuade the reader about a position.
These papers do not test relationships among variables (statistically or otherwise).  

Examining the venues in more detail reveals that the papers published at SIGCSE TS are overwhelmingly \textit{Quantitative} followed by \textit{Survey}.
SIGCSE TS has very few \textit{Qualitative} papers in the years studied.
The ITiCSE papers are also mostly \textit{Quantitative}, but do have a higher percentage of \textit{Qualitative} papers.  
ICER and CSE had the most balanced split between \textit{Quantitative} and \textit{Qualitative} studies, but a smaller percentage of \textit{Survey} papers.

Over two-thirds of empirical studies in CER literature are \textit{Quantitative} (67\%). 
We coded only five papers with multiple evaluation methods.
These papers were either \textit{Survey} -- \textit{Qualitative} or \textit{Survey} -- \textit{Quantitative}.  
While the large emphasis that these CER papers place on \textit{Quantitative} methods provides interesting numerical results, it does suggest the lack of nuance that \textit{Qualitative} methods bring to help understand the \textit{why} and \textit{how} of a study, which can \textit{expand} and \textit{deepen} our understanding of computing phenomena~\cite{bd12,hazzan2006qualitative}.

\begin{table}[!htb]
\centering
\caption{Evaluation Subject by Venue}
\label{tab_evaluation_subjects}
\begin{tabular}{|l||r|r||r|r||r|r||r|r||r|r||r|r|}
\hline
\textbf{Evaluation Subject}                                                   & \multicolumn{2}{c||}{\textbf{\begin{tabular}[c]{@{}c@{}}SIGCSE \\ TS\end{tabular}}} & \multicolumn{2}{c||}{\textbf{ICER}} & \multicolumn{2}{c||}{\textbf{ITiCSE}} & \multicolumn{2}{c||}{\textbf{TOCE}} &
\multicolumn{2}{c||}{\textbf{CSE}} & \multicolumn{2}{c|}{\textbf{TOTAL}}\\ \hline
                                                                 & \textbf{\#}       & \textbf{\%}      & \textbf{\#}      & \textbf{\%}     & \textbf{\#}       & \textbf{\%}      & \textbf{\#}      & \textbf{\%}      & \textbf{\#}      & \textbf{\%} & \textbf{\#}      & \textbf{\%}    \\ \hline \hline
\textbf{Assignment}                                                       & 13                & 7\%            & 4                & 11\%          & 2                 & 2\%            & 2                & 7\%  & 0 & 0\%  & 21 & 6\%         \\ \hline
\textbf{Community}                                                        & 24                & 14\%           & 7                & 18\%          & 15                & 18\%           & 8                & 28\%  & 2 & 9\%  & 56 & 16\%       \\ \hline
\textbf{Curriculum}                                                       & 50                & 28\%           & 3                & 8\%           & 13                & 15\%           & 7                & 24\%  & 8 & 35\% & 81 & 23\%       \\ \hline
\textbf{Pedagogical Technique} & 59                & 34\%           & 17               & 45\%          & 34                & 40\%           & 9                & 31\%    & 7 & 30\% & 126 & 36\%      \\ \hline
\textbf{Tool}                                                             & 24                & 14\%           & 3                & 8\%           & 17                & 20\%           & 2                & 7\%     & 2 & 9\% & 48 & 14\%     \\ \hline
\textbf{Other}                                                            & 6                 & 3\%            & 4                & 11\%          & 4                 & 5\%            & 2                & 7\%     & 4 & 17\%   & 20 & 6\%     \\ \hline
\end{tabular}
\end{table}

\hfill \break
\noindent
\textbf{Evaluation Subjects (Table~\ref{tab_evaluation_subjects}).} \textit{Pedagogical Techniques}, or teaching methods, were the most common evaluation subject in all venues. 
\textit{Tools} papers were more common at the SIGCSE TS and ITiCSE. 
All venues had papers evaluating the broader \textit{Community}.  
\textit{Curriculum} papers were more common at the SIGCSE TS, TOCE, and CSE. 
Readers may find the distribution of paper types at these venues for 2014 and 2015 interesting as they consider future studies that either fit with or fill gaps with existing literature.

\begin{table}[!htb]
\centering
\caption{Evaluation Subject Source by Venue}
\label{tab_evaluation_subject_source}
\begin{tabular}{|l||r|r||r|r||r|r||r|r||r|r||r|r|}
\hline
\textbf{\begin{tabular}[c]{@{}l@{}}Evaluation Subject\\   Source\end{tabular}} & \multicolumn{2}{c||}{\textbf{\begin{tabular}[c]{@{}c@{}}SIGCSE \\ TS\end{tabular}}} & \multicolumn{2}{c||}{\textbf{ICER}} & \multicolumn{2}{c||}{\textbf{ITiCSE}} & \multicolumn{2}{c||}{\textbf{TOCE}} &
\multicolumn{2}{c||}{\textbf{CSE}} &
\multicolumn{2}{c|}{\textbf{TOTAL}} \\ \hline
\textbf{}                                                                      & \textbf{\#}       & \textbf{\%}      & \textbf{\#}      & \textbf{\%}     & \textbf{\#}       & \textbf{\%}      & \textbf{\#}      & \textbf{\%}    & \textbf{\#}      & \textbf{\%} & \textbf{\#}      & \textbf{\%}  \\ \hline \hline
\textbf{Authors Here}                                                                   & 101               & 57\%           & 21               & 55\%          & 51                & 60\%           & 17               & 59\%  & 18 & 82\%  & 208 & 59\%       \\ \hline
\textbf{Authors Elsewhere}                                                              & 11                & 6\%            & 4                & 11\%          & 9                 & 11\%           & 3                & 10\%    & 1 & 5\% & 28 & 8\%      \\ \hline
\textbf{Other Modified}                                                                 & 32                & 18\%           & 6                & 16\%          & 8                 & 9\%            & 1                & 3\%      & 2 & 9\%   & 49 & 14\%    \\ \hline
\textbf{Other Not Modified}                                                             & 10                & 6\%            & 7                & 18\%          & 6                 & 7\%            & 4                & 14\%     & 0 & 0\%   & 27 & 8\%   \\ \hline
\textbf{Community}                                                                      & 22                & 13\%           & 1                & 3\%           & 11                & 13\%           & 4                & 14\%     & 1 & 5\%   & 39 & 11\%    \\ \hline
\end{tabular}
\end{table}

\hfill \break
\noindent
\textbf{Evaluation Subject Source (Table~\ref{tab_evaluation_subject_source}).} 
This rubric item is a measure of replication and reproducibility in the community.  
This item could take one of five values: (1) \textit{Authors Here} -- the authors created the evaluation subject for use in the current study; (2) \textit{Authors Elsewhere} -- the authors created and published the evaluation subject in a previous publication; (3) \textit{Other Modified} -- someone other than the authors created the evaluation subject, but the authors modified it;  (4) \textit{Other Not Modified} -- someone other than the authors created the evaluation subject, and the authors did not modify it; or (5) \textit{Community} -- self-identifying group of people related to an area of interest (if the evaluation subject is \textit{Community}, then the source is \textit{Community} as well).

More than half of all empirical papers used evaluation subjects developed for the specific study (i.e. \textit{Authors Here}), indicating that researchers may be working in isolation and not using existing data sets or possible comparisons to similar participants in a different context.
Papers that reuse an evaluation subject from a previous study by another author demonstrate replication and sharing within the broader community~\cite{NSF19002,schmidt2009replication}.
In some cases, the authors conducted a study on an evaluation source that they utilized in earlier work, which we coded as \textit{Authors Elsewhere}.  
Overall, ICER had the highest percentage of papers in which authors utilized evaluation subjects from other researchers (e.g. \textit{Other Modified} and \textit{Other Not Modified}).

\begin{table}[!htb]
\centering
\caption{Comparison in Evaluation by Venue}
\label{tab_comparison}
\begin{tabular}{|l||r|r||r|r||r|r||r|r||r|r||r|r|}
\hline
\textbf{\begin{tabular}[c]{@{}l@{}}Comparison in \\ Evaluation\end{tabular}} & \multicolumn{2}{c||}{\textbf{\begin{tabular}[c]{@{}c@{}}SIGCSE \\ TS\end{tabular}}} & \multicolumn{2}{c||}{\textbf{ICER}} & \multicolumn{2}{c||}{\textbf{ITiCSE}} & \multicolumn{2}{c||}{\textbf{TOCE}} &
\multicolumn{2}{c||}{\textbf{CSE}} &
\multicolumn{2}{c|}{\textbf{TOTAL}} \\ \hline
\textbf{}           & \textbf{\#}       & \textbf{\%}      & \textbf{\#}      & \textbf{\%}     & \textbf{\#}       & \textbf{\%}      & \textbf{\#}      & \textbf{\%} & \textbf{\#}      & \textbf{\%} & \textbf{\#}      & \textbf{\%}    \\ \hline \hline
\textbf{Comparison} & 87                & 49\%           & 20               & 53\%          & 46                & 54\%           & 15               & 52\% & 8 & 35\%   & 176 & 50\%        \\ \hline
\textbf{None}       & 89                & 51\%           & 18               & 47\%          & 39                & 46\%           & 14               & 48\%    & 15 & 65\%  & 175 & 50\%      \\ \hline
\end{tabular}
\end{table}

\hfill \break
\noindent
\textbf{Comparison (Table~\ref{tab_comparison}).} 
By comparing the results from an intervention to the results from a baseline approach, a paper can provide additional support to strengthen a conclusion that the intervention caused the observed effect. 
The lack of \textit{Comparison} leaves open the possibility of confounding factors influencing the observed result.
However, when papers report a case study (i.e. an in-depth analysis of a single case), the lack of comparison is expected (e.g., \textit{None}). 
Case studies can provide valuable findings that other researchers can attempt to replicate in their own environments, using more experimental approaches.
Approximately half of the papers in each venue did provide some type of \textit{Comparison} between two or more groups of participants.  
To promote comparisons between researchers,~\citet{margulieux2019measurements} suggest utilizing common, standardized, measurements.

\begin{table}[!htb]
\centering
\caption{Number of Participants by Venue}
\label{tab_subjects}
\begin{tabular}{|l||r|r||r|r||r|r||r|r||r|r||r|r|}
\hline
\textbf{\begin{tabular}[c]{@{}l@{}}Number of \\ Participants\end{tabular}} & \multicolumn{2}{c||}{\textbf{\begin{tabular}[c]{@{}c@{}}SIGCSE \\ TS\end{tabular}}} & \multicolumn{2}{c||}{\textbf{ICER}} & \multicolumn{2}{c||}{\textbf{ITiCSE}} & \multicolumn{2}{c||}{\textbf{TOCE}} &
\multicolumn{2}{c||}{\textbf{CSE}} &
\multicolumn{2}{c|}{\textbf{TOTAL}}\\ \hline
\textbf{}                   & \textbf{\#}       & \textbf{\%}      & \textbf{\#}      & \textbf{\%}     & \textbf{\#}       & \textbf{\%}      & \textbf{\#}      & \textbf{\%} & \textbf{\#}      & \textbf{\%}  & \textbf{\#}      & \textbf{\%}   \\ \hline \hline
\textbf{1}                  & 6                 & 3\%            & 0                & 0\%           & 2                 & 2\%            & 1                & 3\%   & 0 & 0\%   & 9 & 3\%       \\ \hline
\textbf{2}                  & 1                 & 1\%            & 2                & 5\%           & 2                 & 2\%            & 1                & 3\% & 0 & 0\%  & 6 & 2\%           \\ \hline
\textbf{3-10}               & 10                & 6\%            & 2                & 5\%           & 6                 & 7\%            & 3                & 10\% & 0 & 0\% & 20 & 6\%          \\ \hline
\textbf{11-30}              & 38                & 22\%           & 7                & 18\%          & 21                & 25\%           & 3                & 10\% & 2 & 10\%  & 71 & 20\%        \\ \hline
\textbf{31-75}              & 38                & 22\%           & 10               & 26\%          & 15                & 18\%           & 6                & 21\%  & 6 & 29\% & 75 & 21\%        \\ \hline
\textbf{76-150}             & 25                & 14\%           & 6                & 16\%          & 13                & 15\%           & 3                & 10\% & 1 & 5\%  & 48 & 14\%       \\ \hline
\textbf{151-300}            & 20                & 11\%           & 5                & 13\%          & 7                 & 8\%            & 3                & 10\%  & 5 & 24\% & 40 & 11\%       \\ \hline
\textbf{301-999}            & 19                & 11\%           & 3                & 8\%           & 12                & 14\%           & 4                & 14\%  & 5 & 24\%   & 43 & 12\%       \\ \hline
\textbf{1000+}              & 13                & 7\%            & 3                & 8\%           & 7                 & 8\%            & 4                & 14\%  & 1 & 5\%   & 21 & 8\%      \\ \hline
\textbf{Missing}            & 6                 & 3\%            & 0                & 0\%           & 0                 & 0\%            & 1                & 3\%   & 1 & 5\%   & 8 & 2\%       \\ \hline
\end{tabular}
\end{table}

\begin{table}[!htb]
\centering
\caption{Summary Statistics of Participants by Venue}
\label{tab_subjects_summary}
\begin{tabular}{|l||r|r|r|r|r||r|}
\hline 
\textbf{Summary Statistics}          & \multicolumn{1}{l|}{\textbf{SIGCSE TS}} & \multicolumn{1}{l|}{\textbf{ICER}} & \multicolumn{1}{l|}{\textbf{ITiCSE}} & \multicolumn{1}{l|}{\textbf{TOCE}} & \multicolumn{1}{l||}{\textbf{CSE}} & \multicolumn{1}{l|}{\textbf{TOTAL}} \\ \hline \hline
Average & 1937.6                         & 913.1                     & 921.2                       & 583.8                     & 314.5                    & 1363.7                     \\ \hline
Median  & 63.5                           & 75                        & 60                          & 95.5                      & 183                      & 72                         \\ \hline
Mode    & 1                              & 75                        & 23                          & 5                         & 183                      & 12                         \\ \hline
\end{tabular}
\end{table}

\hfill \break
\noindent
\textbf{Participants (Table~\ref{tab_subjects} \& Table~\ref{tab_subjects_summary}).} 
The number of \textit{Participants} in published studies can provide some insight into the strength of the conclusions drawn.
Of concern are the papers that did not provide any numbers when discussing the participants in the study. Of the 349 empirical papers that should report subject data\footnote{Two CSE papers categorized as descriptive had N/A recorded for subjects.}, 98\% of papers reported subjects in some form.  This percentage is higher than the 83.7\% of articles reporting subjects in ~\citet{mcgill2018improving}. We additionally report summary statistics for the number of participants by venue and across all venues. Since we are considering all publications in the venues rather than a subset based on an inclusion criteria, the mean and median number of participants is higher than the mean of 328 and median of 45 reported by McGill et al., likely due to the K-12 focus ~\cite{mcgill2018improving}.

\begin{table}[!htb]
\centering
\caption{Type of Study for Non-Descriptive Papers by Venue and for \textit{Quantitative} Only Papers by Venue}
\label{tab_type_study}
\begin{tabular}{|l||r|r||r|r||r|r||r|r||r|r||r|r|}
\hline
 & \multicolumn{2}{c||}{\textbf{\begin{tabular}[c]{@{}c@{}}SIGCSE \\ TS\end{tabular}}} & \multicolumn{2}{c||}{\textbf{ICER}} & \multicolumn{2}{c||}{\textbf{ITiCSE}} & \multicolumn{2}{c||}{\textbf{TOCE}} &
\multicolumn{2}{c||}{\textbf{CSE}} &\multicolumn{2}{c|}{\textbf{TOTAL}} \\ \hline \hline
\textbf{\begin{tabular}[c]{@{}l@{}}Type of Study \\ (Non-Descriptive)\end{tabular}}                                                                                 & \textbf{\#}       & \textbf{\%}      & \textbf{\#}      & \textbf{\%}     & \textbf{\#}       & \textbf{\%}      & \textbf{\#}      & \textbf{\%} & \textbf{\#}      & \textbf{\%} & \textbf{\#}      & \textbf{\%}    \\ \hline 
\textbf{Observational}                                                                    & 86                & 51\%           & 20               & 53\%          & 37                & 44\%           & 16               & 59\% &  12               & 60\% & 171 & 51\%         \\ \hline
\textbf{Interventional}                                                                   & 83                & 49\%           & 18               & 47\%          & 47                & 56\%           & 11               & 41\% &  8               & 40\% & 167 & 49\%         \\ \hline
\hline
\textbf{\begin{tabular}[c]{@{}l@{}}Type of Study  \\ (Quantitative Only)\end{tabular}}    & \textbf{\#}       & \textbf{\%}      & \textbf{\#}      & \textbf{\%}     & \textbf{\#}       & \textbf{\%}      & \textbf{\#}      & \textbf{\%}   & \textbf{\#}      & \textbf{\%}  & \textbf{\#}      & \textbf{\%}   \\ \hline
\textbf{Observational}                                                                    & 63                & 50\%           & 9                & 41\%          & 20                & 34\%           & 8                & 47\% & 5                 & 42\% & 105 & 44\%        \\ \hline
\textbf{Interventional}                                                                   & 63                & 50\%           & 13               & 59\%          & 39                & 66\%           & 9                & 53\%  & 7                & 58\%  & 131 & 56\%        \\ \hline
\end{tabular}
\end{table}

\hfill \break
\noindent
\textbf{Type of Study (Table~\ref{tab_type_study}).} 
An \textit{Observational} study is performed in a natural setting in which the researcher collects data via observation without manipulation of the situation.  
Conversely, an \textit{Interventional} study is performed by assigning participants into groups (e.g., control and experimental) and applying the treatment to the experimental group to measure its effect. Our use of observational and interventional are not intended as a direct relationship to qualitative and quantitative methods, but are instead a way of describing the setting. \textit{Observational} studies are similar to Fincher's and Petre's \cite{fp04} definition on \textit{in situ} -- the normal or natural setting.  \textit{Interventional} studies correspond to Fincher's and Petre's \cite{fp04} settings of \textit{under constraints} and \textit{in a laboratory} where there is some direct intervention to the environment.

There is a fairly even split between the two types of studies across all venues.
This result indicates a nice balance in CER work.  
However, when looking at \textit{Quantitative} work only, we do see an increase in the proportion of \textit{Interventional} studies. 
These results suggest that there is a gap in using \textit{Qualitative} methods to assess CER interventions.

\begin{table}[!htb]
\centering
\caption{Data Source of Papers by Venue}
\label{tab_data_source}
\begin{tabular}{|l|l||r|r|r|r|r|r|r|r||r|r|}
\hline
\multicolumn{2}{|c||}{\textbf{\begin{tabular}[c]{@{}c@{}}Data \\Source\end{tabular}}} & \multicolumn{1}{c|}{\rotatebox[origin=c]{90}{\textbf{\begin{tabular}[c]{@{}c@{}}Assessment \\ Data\end{tabular}}}} & 
\multicolumn{1}{c|}{\rotatebox[origin=c]{90}{\textbf{Automated}}} & \multicolumn{1}{c|}{\rotatebox[origin=c]{90}{\textbf{\begin{tabular}[c]{@{}c@{}}Course \\ Evaluations\end{tabular}}}} & 
\multicolumn{1}{c|}{\rotatebox[origin=c]{90}{\textbf{Interviews}}} & \multicolumn{1}{c|}{\rotatebox[origin=c]{90}{\textbf{\begin{tabular}[c]{@{}c@{}}Focus \\ Group\end{tabular}}}} & 
\multicolumn{1}{c|}{\rotatebox[origin=c]{90}{\textbf{Observations}}} & \multicolumn{1}{c|}{\rotatebox[origin=c]{90}{\textbf{Survey}}} & \multicolumn{1}{c||}{\rotatebox[origin=c]{90}{\textbf{Other}}} & \multicolumn{1}{c|}{\rotatebox[origin=c]{90}{\textbf{\begin{tabular}[c]{@{}c@{}}More than One\end{tabular}}}} & \multicolumn{1}{c|}{\textbf{\rotatebox[origin=c]{90}{\begin{tabular}[c]{@{}c@{}}Survey ONLY\end{tabular}}}} \\ \hline \hline
\multirow{3}{*}{\rotatebox[origin=c]{90}{\textbf{SIGCSE}}} & \textbf{Quant}  & 46\%                                                                                     & 13\%                                    & 7\%                                                                                         & 6\%                                      & 1\%                                                                                  & 8\%                                        & 35\%                                 & 5\%                                 & 19\%                                                                                      & 24\%                                                                                \\ \cline{2-12} 
                                 & \textbf{Qual} & 40\%                                                                                     & 0\%                                     & 0\%                                                                                         & 47\%                                     & 13\%                                                                                 & 7\%                                        & 27\%                                 & 0\%                                 & 33\%                                                                                      & 7\%                                                                                 \\ \cline{2-12} 
                                 & \textbf{Sur}  & 6\%                                                                                      & 0\%                                     & 0\%                                                                                         & 3\%                                      & 3\%                                                                                  & 0\%                                        & 100\%                                & 0\%                                 & 13\%                                                                                      & 88\%                                                                                \\ \hline \hline
\multirow{3}{*}{\rotatebox[origin=c]{90}{\textbf{ICER}}}   & \textbf{Quant}  & 45\%                                                                                     & 27\%                                    & 0\%                                                                                         & 0\%                                      & 0\%                                                                                  & 5\%                                        & 32\%                                 & 14\%                                & 23\%                                                                                      & 14\%                                                                                \\ \cline{2-12} 
                                 & \textbf{Qual} & 8\%                                                                                      & 8\%                                     & 0\%                                                                                         & 67\%                                     & 0\%                                                                                  & 33\%                                       & 0\%                                  & 17\%                                & 25\%                                                                                      & 0\%                                                                                 \\ \cline{2-12} 
                                 & \textbf{Sur}  & 0\%                                                                                      & 0\%                                     & 0\%                                                                                         & 0\%                                      & 25\%                                                                                 & 0\%                                        & 100\%                                & 0\%                                 & 25\%                                                                                      & 75\%                                                                                \\ \hline \hline
\multirow{3}{*}{\rotatebox[origin=c]{90}{\textbf{ITiCSE}}} & \textbf{Quant}  & 53\%                                                                                     & 17\%                                    & 2\%                                                                                         & 5\%                                      & 0\%                                                                                  & 12\%                                       & 47\%                                 & 10\%                                & 41\%                                                                                      & 15\%                                                                                \\ \cline{2-12} 
                                 & \textbf{Qual} & 23\%                                                                                     & 15\%                                    & 0\%                                                                                         & 15\%                                     & 8\%                                                                                  & 15\%                                       & 15\%                                 & 15\%                                & 8\%                                                                                       & 8\%                                                                                 \\ \cline{2-12} 
                                 & \textbf{Sur}  & 0\%                                                                                      & 0\%                                     & 0\%                                                                                         & 0\%                                      & 0\%                                                                                  & 0\%                                        & 100\%                                & 0\%                                 & 0\%                                                                                       & 100\%                                                                               \\ \hline \hline
\multirow{3}{*}{\rotatebox[origin=c]{90}{\textbf{TOCE}}}   & \textbf{Quant}  & 47\%                                                                                     & 12\%                                    & 0\%                                                                                         & 12\%                                     & 6\%                                                                                  & 12\%                                       & 29\%                                 & 6\%                                 & 24\%                                                                                      & 18\%                                                                                \\ \cline{2-12} 
                                 & \textbf{Qual} & 0\%                                                                                      & 0\%                                     & 0\%                                                                                         & 50\%                                     & 0\%                                                                                  & 25\%                                       & 0\%                                  & 25\%                                & 0\%                                                                                       & 0\%                                                                                 \\ \cline{2-12} 
                                 & \textbf{Sur}  & 14\%                                                                                     & 14\%                                    & 0\%                                                                                         & 0\%                                      & 0\%                                                                                  & 0\%                                        & 100\%                                & 0\%                                 & 29\%                                                                                      & 71\%                                                                                \\ \hline \hline
\multirow{3}{*}{\rotatebox[origin=c]{90}{\textbf{CSE}}}    & \textbf{Quant}  & 67\%                                                                                     & 0\%                                     & 0\%                                                                                         & 25\%                                     & 0\%                                                                                  & 8\%                                        & 58\%                                  & 8\%                                & 42\%                                                                                      & 25\%                                                                                \\ \cline{2-12} 
                                 & \textbf{Qual} & 20\%                                                                                     & 0\%                                     & 0\%                                                                                         & 0\%                                      & 20\%                                                                                 & 40\%                                       & 40\%                                 & 20\%                                & 40\%                                                                                      & 40\%                                                                                \\ \cline{2-12} 
                                 & \textbf{Sur}  & 0\%                                                                                      & 0\%                                     & 0\%                                                                                         & 0\%                                      & 25\%                                                                                 & 0\%                                        & 100\%                                  & 0\%                               & 25\%                                                                                      & 25\%                                                                                \\ \hline \hline
\multicolumn{2}{|l||}{\textbf{Totals}}                     & 38\%                                                                                     & 11\%                                    & 3\%                                                                                         & 11\%                                     & 3\%                                                                                  & 9\%                                        & 47\%                                 & 7\%                                & 24\%                                                                                      & 30\%                                                                                \\ \hline
\end{tabular}
\end{table}

\hfill \break
\noindent
\textbf{Data Source (Table \ref{tab_data_source}).}
We identified seven key data sources and included an option for data sources that do not fit any of the existing categories.  
Overall, the most common data sources utilized in \textit{Quantitative}, \textit{Qualitative}, and \textit{Survey} work were \textit{Surveys}, \textit{Assessment Data}, and \textit{Automated Data}. 
In addition, the \textit{Qualitative} studies frequently featured \textit{Interviews}, \textit{Focus Groups}, and \textit{Observations}.  Our categorization of data sources has some overlap with related work, but our focus was identifying high-level data sources.  That means that categories in other studies, like attitudinal data described in ~\citet{randolph2008methodological} was typically grouped in a larger category of ``Survey''.   

The analysis showed that 76\% of the papers 
utilized only one type of data source, even if they collected multiple samples. 
By only utilizing a single data source, authors limit their understanding of the phenomena under study because they cannot triangulate the results to clarify or understand~\cite{fp04,margulieux2019measurements}.
The analysis also showed that 30\% of all papers used \textit{Surveys} as the sole data source. 
While surveys are an excellent tool for collecting attitudinal and self-report data, as the only source of data, they may not be robust enough to provide strong evidence to answer a research question.  We did not collect data on the use of validated survey instruments as described in~\citet{decker-sigcsets2019-instruments}.

\subsection{RQ3: Norms of Empirical Evaluation}

\begin{tcolorbox}[enhanced,drop shadow]
    \textbf{RQ3: Of the papers that have empirical evaluation, do they follow norms (both inclusion and labeling of information needed for replication, meta-analysis, and, eventually, theory-building) for reporting empirical work?}
\end{tcolorbox}

For empirical papers, we expect a paper should contain a research objective, related work, an overview of the research participants, an overview of the study design or methods, a description of the data collection process, a description of the analysis procedures, a report of the results, and a listing of the threats to validity for the research~\cite{fp04,bd12,creswell_creswell_2018,Schulzc-BMJ2010-ConsortStatement} as defined in the \textit{CER Empiricism Assessment -- Experimental Rubric}.  
Any papers reporting the results of a \textit{Survey} as a data source should have information about how the survey was conducted and the survey questions asked as defined in the \textit{CER Empiricism Assessment -- Survey Rubric}.
\textit{Descriptive/Persuasive} work should include the goal of the argument, related work, at least three premises and a conclusion, all with supporting evidence as defined in the \textit{CER Empiricism Assessment -- Descriptive/Persuasive Rubric}. However, due to the small numbers of \textit{Descriptive/Persuasive} papers in our analysis (e.g., 10), we do not provide an analysis of that category in this section.

To provide an overall characterization of the papers, we classified each paper into one of three categories for each of the key elements listed above:
\begin{itemize}
    \item \textbf{Strongly Supports Replication} - items that were ``Complete and Labeled'' provide all the details necessary to support replication;
    \item \textbf{Weakly Supports Replication} - items that were ``Complete and Not Labeled'' or ``Partial and Labeled'' provide some, but not all of the details required to support replication;
    \item \textbf{Not Present} - items that were ``Not Present'' are missing information necessary to support replication.
\end{itemize}

Table~\ref{tab_norms} summarizes the results of our analysis of reporting norms.  
Overall, the only two elements where more than 50\% of papers reach the \textbf{Strongly Supports Replication} level are \textit{Related Work} and \textit{Results}.
Looking across all elements, the papers published in ICER do a better job by having more then 50\% of papers reach the \textbf{Strongly Supports Replication} level for all elements except for Threats to Validity.
However, as the numbers in Table~\ref{tab_norms} show, there is still room for improvement as a community to reach the \textbf{Strongly Supports Replication} level for all attributes most (or all) of the time.

\subsubsection{Experimental Rubric}

\begin{table}[!htb]
\centering
\caption{Norms of Reporting Empirical Evaluation}
\label{tab_norms}
\begin{tabular}{|l|l||r|r|r|r|r||r|}
\hline
                                                                                        &                   \multicolumn{1}{c||}{\textbf{\begin{tabular}[c]{@{}l@{}}Replication\\ Support\end{tabular}}}          & \multicolumn{1}{c|}{\textbf{\begin{tabular}[c]{@{}c@{}}SIGCSE \\ TS\end{tabular}}} & \multicolumn{1}{c|}{\textbf{ICER}} & \multicolumn{1}{c|}{\textbf{ITiCSE}} & \multicolumn{1}{c|}{\textbf{TOCE}} & \multicolumn{1}{c||}{\textbf{CSE}} & \multicolumn{1}{c|}{\textbf{TOTAL}} \\ \hline \hline
\multirow{3}{*}{\textbf{\begin{tabular}[c]{@{}l@{}}Research\\ Objective\end{tabular}}}  & \textbf{Strong} & 20.1\%                               & 55.3\%                             & 17.9\%                               & 55.6\%                             & 55.0\%                            & 28.4\%                              \\ \cline{2-8} 
                                                                                        & \textbf{Weak}   & 58.0\%                               & 31.6\%                             & 61.9\%                               & 37.0\%                             & 45.0\%                            & 53.6\%                              \\ \cline{2-8} 
                                                                                        & \textbf{Not Present}        & 21.9\%                               & 13.2\%                             & 20.2\%                               & 7.4\%                              & 0.0\%                             & 18.0\%                              \\ \hline \hline
\multirow{3}{*}{\textbf{\begin{tabular}[c]{@{}l@{}}Related\\ Work\end{tabular}}}        & \textbf{Strong} & 62.1\%                               & 81.6\%                             & 70.2\%                               & 81.5\%                             & 65.0\%                            & 68.0\%                              \\ \cline{2-8} 
                                                                                        & \textbf{Weak}   & 27.8\%                               & 15.8\%                             & 19.0\%                               & 18.5\%                             & 30.0\%                            & 23.7\%                              \\ \cline{2-8} 
                                                                                        & \textbf{Not Present}        & 10.1\%                               & 2.6\%                              & 10.7\%                               & 0.0\%                              & 5.0\%                             & 8.3\%                               \\ \hline \hline
\multirow{3}{*}{\textbf{Subjects}}                                                      & \textbf{Strong} & 32.0\%                               & 50.0\%                             & 27.4\%                               & 44.4\%                             & 44.4\%                            & 34.5\%                              \\ \cline{2-8} 
                                                                                        & \textbf{Weak}   & 60.4\%                               & 47.4\%                             & 61.9\%                               & 51.9\%                             & 50.0\%                            & 58.0\%                              \\ \cline{2-8} 
                                                                                        & \textbf{Not Present}        & 7.7\%                                & 2.6\%                              & 10.7\%                               & 3.7\%                              & 5.6\%                             & 7.4\%                               \\ \hline \hline
\multirow{3}{*}{\textbf{\begin{tabular}[c]{@{}l@{}}Study\\ Design\end{tabular}}}        & \textbf{Strong} & 33.7\%                               & 60.5\%                             & 56.0\%                               & 48.1\%                             & 68.8\%                            & 45.2\%                              \\ \cline{2-8} 
                                                                                        & \textbf{Weak}   & 59.2\%                               & 36.8\%                             & 39.3\%                               & 51.9\%                             & 31.3\%                            & 49.7\%                              \\ \cline{2-8} 
                                                                                        & \textbf{Not Present}        & 7.1\%                                & 2.6\%                              & 4.8\%                                & 0.0\%                              & 0.0\%                             & 5.1\%                               \\ \hline \hline
\multirow{3}{*}{\textbf{\begin{tabular}[c]{@{}l@{}}Data\\ Collection\end{tabular}}}     & \textbf{Strong} & 27.8\%                               & 52.6\%                             & 23.8\%                               & 25.9\%                             & 60.0\%                            & 30.9\%                              \\ \cline{2-8} 
                                                                                        & \textbf{Weak}   & 65.7\%                               & 44.7\%                             & 70.2\%                               & 66.7\%                             & 40.0\%                            & 63.4\%                              \\ \cline{2-8} 
                                                                                        & \textbf{Not Present}        & 6.5\%                                & 2.6\%                              & 6.0\%                                & 7.4\%                              & 0.0\%                             & 5.7\%                               \\ \hline \hline
\multirow{3}{*}{\textbf{\begin{tabular}[c]{@{}l@{}}Analysis\\ Procedures\end{tabular}}} & \textbf{Strong} & 28.6\%                               & 57.9\%                             & 23.8\%                               & 48.1\%                             & 71.4\%                            & 34.1\%                              \\ \cline{2-8} 
                                                                                        & \textbf{Weak}   & 60.7\%                               & 39.5\%                             & 65.5\%                               & 51.9\%                             & 28.6\%                            & 57.4\%                              \\ \cline{2-8} 
                                                                                        & \textbf{Not Present}        & 10.7\%                               & 2.6\%                              & 10.7\%                               & 0.0\%                              & 0.0\%                             & 8.5\%                               \\ \hline \hline
\multirow{3}{*}{\textbf{Results}}                                                       & \textbf{Strong} & 61.5\%                               & 97.4\%                             & 78.6\%                               & 81.5\%                             & 100.0\%                           & 73.5\%                              \\ \cline{2-8} 
                                                                                        & \textbf{Weak}   & 37.3\%                               & 2.6\%                              & 21.4\%                               & 18.5\%                             & 0.0\%                             & 25.9\%                              \\ \cline{2-8} 
                                                                                        & \textbf{Not Present}        & 1.2\%                                & 0.0\%                              & 0.0\%                                & 0.0\%                              & 0.0\%                             & 0.6\%                               \\ \hline \hline
\multirow{3}{*}{\textbf{Threats}}                                                       & \textbf{Strong} & 7.7\%                                & 31.6\%                             & 4.8\%                                & 25.9\%                             & 20.0\%                            & 11.8\%                              \\ \cline{2-8} 
                                                                                        & \textbf{Weak}   & 13.0\%                               & 23.7\%                             & 13.1\%                               & 7.4\%                              & 20.0\%                            & 14.2\%                              \\ \cline{2-8} 
                                                                                        & \textbf{Not Present}        & 79.3\%                               & 44.7\%                             & 82.1\%                               & 66.7\%                             & 60.0\%                            & 74.0\%                              \\ \hline
\end{tabular}
\end{table}

The \textit{Experimental Rubric} describes the standard information we would expect to see in a paper reporting on empirical work. 
Table~\ref{tab_norms} summarizes the results for the items on the \textit{Experimental Rubric}.

\hfill \break
\noindent
\textbf{Research Question.} One of the most critical pieces of information in an empirical paper is the research objective, goals, hypotheses, or other summative statement that provides the context of the work reported in the paper.  
It is concerning that, across all venues, 18\% of the papers lacked a summative statement about the work (i.e. scored as \textbf{Not Present}).  \citet{randolph2008methodological} reported only 22\% of papers between 2000 and 2005 had research questions or hypotheses and \citet{lishinski2016methodological} found that only 47\% of CSE and 56\% of ICER papers between 2012 and 2015 had explicit research questions. Our numbers for CSE (55\%) and ICER (55.3\%) for complete and labeled are similar to \citet{lishinski2016methodological} and higher than \citet{randolph2008methodological} suggesting some improvement in the reporting of research questions.

A summative statement guides the reader to the main ideas of the study and is critical for determining if the methods and analysis are appropriate for the statement and if the results answer or address the statement.  One or more research statements, objectives, or hypotheses are recommended in standards like CONSORT~\cite{Schulzc-BMJ2010-ConsortStatement}, APA JARS~\cite{apa-jars}, and AERA~\cite{area-empirical-social-science-2006}. 
In addition, we found over half of the papers reached only the \textbf{Weakly Supports Replication} level because they lacked important information.
These papers either did not highlight the summative statement in a meaningful way, through text attributes (e.g., italics or bold), callouts (e.g., boxes or bullets), or clear labeling (e.g., RQ1, Goal, etc.), or did not clearly label the goal of the paper, thereby requiring the reader to infer the context of the paper through other statements in the paper.

\hfill \break
\noindent
\textbf{Related Work.} Most papers reported and labeled related work demonstrating \textbf{Strongly Supports Replication} for this category. 
The SIGCSE Board policy\footnote{See SIGCSE Board Program Chair Responsibilities policy at \url{https://sigcse.org/policies/pcr.html}.} states that Program Chairs should ``request that all paper submissions include a review of previous, related work.'' 
This guidance demonstrates the importance of establishing reporting norms for the community to ensure high quality dissemination of empirical results.
The papers we rated as \textbf{Weakly Supports Replication} typically did not link the related work to the research objective or did not contain an explicitly labeled related work or background section. 
For those papers that lacked a related work section, the authors typically discussed the related work or background information in the introduction. 

\hfill \break
\noindent
\textbf{Participants.} Most papers did include information about the study participants.
However, over half of the papers fell into the \textbf{Weakly Supports Replication} category because they lacked either full details about the participants or a clear label for the section of the paper describing the participants. 
A description of participants should contain demographic information to provide context for the results~\cite{mcgill2018improving,NSF19002,apa-jars,Schulzc-BMJ2010-ConsortStatement,area-empirical-social-science-2006}.  
For example, when papers include student populations, the demographics should include: the number of participants, age groups, education levels, gender, race/ethnicity, region or area of the world, and prior computing experience~\cite{mcgill2018improving}. 
Other student demographics that may be of interest depending on the study include student disabilities, socioeconomic status, family history (e.g., first generation college student), and veteran status.  
Demographics may vary based on the populations under study and the research questions.

In addition to lacking demographics, papers often lacked a description of the recruitment process for the participants.
Classroom research may lack the formal control and treatments present in laboratory or controlled studies. 
By clarifying the inclusion and exclusion criteria for participants, an author clarifies the participant pool and research context~\cite{NSF19002,Schulzc-BMJ2010-ConsortStatement,apa-jars,area-empirical-social-science-2006}. 
Papers should also include a statement about how the authors recruited participants and obtained consent to participate in the study.
This statement demonstrates ethical treatment of research participants and assure readers that authors are following standards for human subject research in their context.

\hfill \break
\noindent
\textbf{Study Design.} Most papers contained a discussion about study design. 
We found a split between \textbf{Strongly Supports Replication} and \textbf{Weakly Supports Replication}.
There were a number of papers that lacked a clear study design or methods section. 
In many of these cases, the study design information may have been integrated with the results, rather than as a separate independent section. 
By providing the study design information in a separate section, the authors help readers more easily extract important information about the design to support replication. 
In addition to not being labeled, some study designs sections omitted key information. 
One key piece of information is identification of the dependent and independent variables~\cite{Schulzc-BMJ2010-ConsortStatement,apa-jars}, which could be as simple as identifying the key item under observation or providing overview of an intervention. 

\hfill \break
\noindent
\textbf{Data Collection.} Most papers contained a discussion about data collection. 
However, 63.4\% of papers fell into the \textbf{Weakly Supports Replication} category.
Some of those papers did not include a clearly defined data collection section. 
Sometimes, the papers integrated the discussion about data collection into the discussion about the study design or methods.  
Other times, papers integrated the discussion about data collection into the results section. 
By providing a clear section or subsection around data collection~\cite{apa-jars,area-empirical-social-science-2006}, authors can provide details to readers who may be interested in using similar techniques in their own work.

Other  \textbf{Weakly Supports Replication} papers omitted important information describing the \textit{how}, \textit{where}, and \textit{who} of the data collection.
For example, papers frequently lacked details about survey administration, which is discussed further in Section~\ref{survey-rubric-text}.

\hfill \break
\noindent
\textbf{Analysis Procedures.} Around 34\% of papers demonstrated \textbf{Strongly Supports Replication} related to the reporting of analysis procedures. 
Around half of the \textbf{Weakly Supports Replication} papers were both incomplete and not properly labeled. 
These papers were commonly missing a full description of how the authors worked with or processed that data after collection. 
Additionally, the papers often did not provide a justification for why the particular statistical tests were appropriate for the analysis.  
A discussion on the use of inferential statistics for ICER's proceedings history provides a justification for full reporting of statistical tests and examples of papers with \textbf{Strongly Supports Replication} in this space~\cite{sanders2019icer}. Standards support clear reporting on analysis procedures and statistical tests used~\cite{apa-jars,area-empirical-social-science-2006}

\hfill \break
\noindent
\textbf{Results.} Most papers reported results and tied the results back to the research question, goal, or hypothesis (which may have been inferred if not explicitly stated).  
A large percentage of ICER and TOCE reached the level of \textbf{Strongly Supports Replication}.   
Papers with \textbf{Weakly Supports Replication} tended to have partial reporting of results, usually by not tying the results back to the research questions, goals, or hypotheses as suggested by APA JARS~\cite{apa-jars}. 
Some papers also omitted a clearly labeled results section. Standards provide guidance on the details that are included when reporting results~\cite{Schulzc-BMJ2010-ConsortStatement,apa-jars, area-empirical-social-science-2006}.

\hfill \break
\noindent
\textbf{Threats.} Most papers (almost 75\%) did not report any threats to validity or limitations to work. Our results are similar to the upper range reported by ~\citet{al2016updated}.  ~\citet{ihantola2015educational} found only 22\% of papers they reviewed on educational data mining reported threats to validity.
If papers did discuss threats or limitations, the threats/limitations were typically unlabeled and appeared in either a results or discussion section.  
Papers with a \textbf{Weakly Supports Replication} may have listed the threats, but typically did not discuss how the authors addressed the threats.
A threats to validity section provides context to the study design and provides details on how readers should interpret the work in a broader context and potential biases that might impact the result~\cite{al2016updated,Schulzc-BMJ2010-ConsortStatement,apa-jars}. 
Because CER is complex work, threats to validity describe how the authors controlled that complexity (or not)~\cite{ihantola2015educational,al2016updated}.  Replications help control for sampling error and artifacts which identify potential weaknesses with internal validity of the original study. Clearly articulated threats to validity can help identify key variables to vary in replications~\cite{schmidt2009replication}.  Replications also support increasing external validity through generalizability~\cite{schmidt2009replication}.

\subsubsection{Survey Rubric}
\label{survey-rubric-text}

We completed the \textit{Survey Rubric} for any paper that contained a survey as a data source.  
To provide replicable survey research, authors must provide details on the survey design and survey execution. 
The reuse and standardization of surveys would be a good starting point for building a culture of replication~\cite{margulieux2019measurements,area-empirical-social-science-2006}.
If authors provide details about their surveys, even as an un-reviewed supplement to the the paper, it would help build this culture.
Table~\ref{tab_norms_survey} summarizes the results for the items on the Survey Rubric.

\begin{table}[!htb]
\centering
\caption{Norms of Reporting Surveys}
\label{tab_norms_survey}
\begin{tabular}{|l|l||r|r|r|r|r||r|}
\hline
& & \multicolumn{1}{l|}{\textbf{\begin{tabular}[c]{@{}c@{}}SIGCSE \\ TS\end{tabular}}} & \multicolumn{1}{l|}{\textbf{ICER}} & \multicolumn{1}{l|}{\textbf{ITiCSE}} & \multicolumn{1}{l|}{\textbf{TOCE}} &
\multicolumn{1}{l||}{\textbf{CSE}} & \multicolumn{1}{l|}{\textbf{TOTAL}}\\ \hline \hline
\multirow{3}{*}{\textbf{\begin{tabular}[c]{@{}l@{}}Conducting \\ the Survey\end{tabular}}} & 
\textbf{Strong Replication} & 28\% & 27\% & 14\% & 18\% & 36\% & 24\% \\ 
\cline{2-8} 
& \textbf{Weak Replication} & 58\% & 73\% & 79\% & 82\% & 64\% & 67\% \\ 
\cline{2-8} 
& \textbf{Not Present} & 14\% & 0\% & 7\% & 0\% & 0\% & 9\% \\ 
\hline \hline
\multirow{3}{*}{\textbf{\begin{tabular}[c]{@{}l@{}}Survey \\ Design\end{tabular}}}     & 
\textbf{Strong Replication} & 30\% & 0\% & 24\% & 36\% & 20\% &  26\% \\ 
\cline{2-8} 
& \textbf{Weak Replication} & 50\% & 91\% & 50\% & 45\% & 20\% & 51\% \\ 
\cline{2-8} 
& \textbf{Not Present} & 20\% & 9\% & 26\% & 18\% & 60\% & 23\% \\ 
\hline
\end{tabular}
\end{table}

\hfill \break
\noindent
\textbf{Conducting the Survey.} We found that most papers demonstrated \textbf{Weak Replication} because they did not fully discuss survey execution in their context. 
The most frequently missing information was details about survey administration and survey medium.  
Additionally, many authors did not clearly label the information about conducting the survey. 

\hfill \break
\noindent
\textbf{Survey Design.} Most papers did not provide a justification for the selection of survey questions based upon how those questions measure items of interest to answer the research questions, which is a category of reporting in APA JARS~\cite{apa-jars,area-empirical-social-science-2006}.  
Additionally, most papers did not include the survey questions, which makes it difficult for others to adopt or reuse the survey in their context. 
In particular, we found that the majority of papers published in CSE during 2014 and 2015 did not discuss the design of the surveys or how the questions were derived, but instead focused more on the distribution and execution of the survey.
Validation and adoption of standardized, citable survey instruments will also support replication~\cite{margulieux2019measurements}.

\section{Discussion}
\label{discussion}
The CER community has seen a growth in submissions and participation. 
As our community grows we need to mature the norms of reporting empirical research so the broader community can benefit from the dissemination of high-quality results, that answer well-stated research questions, situated in the appropriate literature, with a clear discussion of limitations and threats to validity.  
Papers should clearly document methods and analysis procedures to allow others to replicate studies in their own contexts. 
Reviewers should begin to expect paper authors to follow reporting norms and use those norms as guidance when performing paper reviews.

In the following sections, we provide observations and recommendations from our review of reporting norms.  Where appropriate, we connect our recommendations to the broader literature and guidelines, particularly APA JARS~\cite{apa-jars}, CONSORT,  ~\cite{Schulzc-BMJ2010-ConsortStatement}, \textit{What Works Clearinghouse}~\cite{WWC2020}, AERA~\cite{area-empirical-social-science-2006}, and the IES/NSF Guidelines~\cite{NSF13126,NSF19002}.

\subsection{Observations and Recommendations from the Results}
Based on the results described in the previous section, we make a number of observations and recommendations for authors. We note that these recommendations may not apply in all situations and that the individual research questions and researcher context will impact the decisions that are made during a study.  

\hfill \break
\noindent
\textbf{Evaluation Methods.}
The results showed that only five papers had used multiple evaluation methods.
We encourage researchers to consider using mixed methods approaches more often in study designs to utilize the complementary strengths of both \textit{Qualitative} and \textit{Quantitative} methods (see \citet{fp04}, \citet{bd12}, APA JARS~\cite{apa-jars} for details on mixed method studies).
However, we acknowledge that it is not always possible or advisable to use mixed methods approaches and that page limits often constrain the level of detail researchers can include in their publications.

\hfill \break
\noindent
\textbf{Evaluation Subject Source.}
The results showed that more than half of the papers used evaluation subjects developed for the specific study.
This result suggests that replication in CER is weak.  
Replication provides a mechanism for generalizing findings about a research question or set of research questions that can lead to the development of theory to support the computing education community~\cite{schmidt2009replication,ihantola2015educational}.  
We challenge the computing education community to consider using common methods, evaluation metrics, and data collection and reporting procedures to support the comparison and aggregation of data to generalize CER findings.  Utilizing standards can support this goal~\cite{Schulzc-BMJ2010-ConsortStatement,WWC2020,NSF19002,apa-jars,area-empirical-social-science-2006}.

\hfill \break
\noindent
\textbf{Data Source.}
The results showed that a large number of papers utilized only one source of data.
We encourage researchers to utilize multiple data sources to provide more robust insight to the phenomena under study and stronger answers to research questions.
In addition, 29\% of the papers used surveys as the only data source.
There are many situations where surveys are an excellent form of data collection (i.e., understanding a community); however, in classroom studies, researchers should supplement surveys with additional information to provide stronger evidence to answer a research question.  
For example, a research question about how an intervention impacts student learning should not rely solely on student's self-report of their learning.  \citet{margulieux2019measurements} argues that collecting both process (e.g., progress or experience) and product (e.g., performance or outcome) data can increase the applicability of the results to a larger group of educators.

\hfill \break
\noindent
\textbf{Study Design.}
We observed a number of papers that lacked a clear study design or methods section.
A study design section should tie the observation or intervention to the research question(s) and provide justification that the observation or intervention will support answering the research question(s). 
The study design section should clearly describe the steps of the study so others could replicate the study in their own environments. 
For subjective measures, the paper should contain a discussion of the coding process and the method for obtaining agreement or consensus among multiple raters. 
These items are important for determining how to properly interpret the results. 
In addition, citing seminal work and standards about the study methods strengthens the methods discussion and can help build the community's knowledge about CER methods~\cite{fp04,bd12,Schulzc-BMJ2010-ConsortStatement,apa-jars,WWC2020,area-empirical-social-science-2006}.

\hfill \break
\noindent
\textbf{Survey Design.}
We observed that most authors did not provide a justification for the specific survey questions included.
While page restrictions can make it difficult to provide the full survey, authors can include the full survey as an un-reviewed supplemental content or on a website.  
In cases where an author does not want to make the survey publicly available because it may bias future results, the author can include a statement to that effect in the paper and provide contact information for interested researchers to obtain a copy of the survey. The use of standardized survey instruments can support replication~\cite{margulieux2019measurements}.

\subsection{Author Guidelines for Empirical Papers}

Many CER researchers may not have any formal training in conducting human subject research in educational environments.  
Examples of high-quality literature and study frameworks for replication can help new researchers get started investigating interesting challenges in their own classrooms, departments, and communities. 
We therefore propose the following guidelines for \textit{CER authors} when preparing their studies and writing up their work.  These guidelines are similar to expectations when designing CER studies~\cite{fp04,bd12} and have elements in common with standards~\cite{Schulzc-BMJ2010-ConsortStatement,WWC2020,NSF19002,apa-jars,area-empirical-social-science-2006}.  CER Venues, like TOCE, are now recommending that authors utilize APA JARS for future submissions.

\vspace{8pt}
\noindent
\textbf{General Reporting Guidelines}
\begin{itemize}
    \item Clearly identify the evaluation method(s) and subject(s) under evaluation in the study~\cite{Schulzc-BMJ2010-ConsortStatement,apa-jars,area-empirical-social-science-2006}.
    \item Clearly identify what work is new to the study and where the study builds on prior work or data~\cite{apa-jars,NSF19002,area-empirical-social-science-2006}.
    \item Clearly identify any comparisons with prior work or with a control group, if appropriate~\cite{apa-jars,NSF19002,area-empirical-social-science-2006}.
    \item Provide information about participants, including demographics~\cite{mcgill2018improving,Schulzc-BMJ2010-ConsortStatement,apa-jars,area-empirical-social-science-2006}.
    \item Identify the study as observational or interventional~\cite{apa-jars}.
    \item Identify data sources~\cite{apa-jars,area-empirical-social-science-2006}.
    \item For surveys, describe the administration process and provide the survey questions.
    \item Where possible, provide supplemental resources that would help others replicate the work, e.g. code used to analyze data on a public version control system or anonymized, aggregate data on a website~\cite{Schulzc-BMJ2010-ConsortStatement,NSF19002}.
    \item Utilize pre-registration and open science to strengthen research integrity and transparency~\cite{NSF19002,Schulzc-BMJ2010-ConsortStatement,mcgill-koli2019-replication,nosek2014method}.
\end{itemize}

\vspace{8pt}
\noindent
\textbf{Research Question}~\cite{Schulzc-BMJ2010-ConsortStatement,apa-jars,area-empirical-social-science-2006}
\begin{itemize}
    \item The research goal or question is of upmost importance because it drives the rest of the work. Utilize resources on writing high quality and actionable research questions.
    \item Highlight the research goal and questions in the text.  For example, italicize the goal statements in the paper's introduction.  List and name more specific research questions (e.g. RQ1, RQ2) in the introduction and revisit them in the results.
\end{itemize}

\vspace{8pt}
\noindent
\textbf{Related Work}
\begin{itemize}
    \item Report related work in its own section. It should situate the research question or goal in the context of the broader literature. Because \textit{most} SIGCSE conferences now provide extra pages for references, authors should be able to provide a robust discussion of related work with sufficient citations of relevant literature.
    \item Provide a discussion that links the related work to the research goal, objectives, or questions for the paper. The section should also discuss how the related work informs the study and how the study builds on previous work. The information about previous work is especially important for replication papers~\cite{apa-jars,area-empirical-social-science-2006}.
    \item Synthesize the key themes in the prior work to provide context and motivation for the current study, which should not simply be an annotated bibliography.
\end{itemize}

\vspace{8pt}
\noindent
\textbf{Participants}
\begin{itemize}
    \item Report demographic information about the participant groups in the study. For students, minimally report numbers, ages, education levels, gender, race/ethnicity, prior computing experience, and regional location for the students~\cite{mcgill2018improving,apa-jars,area-empirical-social-science-2006}. ~\citet{mcgill2018improving} provide other suggestions for reporting demographic information for other participant populations.  
    \item If number of participants is too small that reporting demographics would become identifying for those participants, clearly state this information in lieu of reporting specific demographics.
    \item Describe the process for selecting participants. If the study uses multiple groups, the paper should discuss the process for allocating participants to each group. If the authors exclude participants from the study (e.g., minors in a university-level study), the paper should explain the exclusion criteria~\cite{Schulzc-BMJ2010-ConsortStatement,apa-jars,area-empirical-social-science-2006}.
    \item Explain the process for obtaining consent from participants and any ethical considerations associated with human subjects research.  Since ethical standards vary by country and institution, authors should clarify expectations in their context and assure reviewers and readers that the human subjects study was handled in an appropriate manner for the author's context~\cite{apa-jars,area-empirical-social-science-2006}.
\end{itemize}

\vspace{8pt}
\noindent
\textbf{Study Design}
\begin{itemize}
    \item Separate the study design or methods from the discussion of results.  This separation will allow readers to assess the quality of the study and to more easily see the steps needed for replication.
    \item Identify independent and dependent variables~\cite{Schulzc-BMJ2010-ConsortStatement,apa-jars}.
    \item Describe each step in conducting the study~\cite{Schulzc-BMJ2010-ConsortStatement,apa-jars,area-empirical-social-science-2006}.
\end{itemize}

\vspace{8pt}
\noindent
\textbf{Data Collection}
\begin{itemize}
    \item Identify data collection procedures including mediums for data collection, who collected the data, and how the data addresses or answers the research questions~\cite{NSF19002,apa-jars,area-empirical-social-science-2006}.
    \item Provide a rationale for why the data collected will help address the research question(s)~\cite{apa-jars,area-empirical-social-science-2006}.
\end{itemize}

\vspace{8pt}
\noindent
\textbf{Analysis Procedures}
\begin{itemize}
    \item Describe the process for working with the collected data, including the process for cleaning data, if necessary~\cite{NSF19002,area-empirical-social-science-2006}.
    \item For qualitative analysis, describe the coding process and the process for evaluating the correctness of the coding (e.g., multiple raters, inter-rater reliability, etc.)~\cite{apa-jars,area-empirical-social-science-2006}.
    \item For quantitative analysis, describe, and justify, the statistical tests, if appropriate.  For other types of analysis, provide a justification on how the analysis helps answer the research question(s)~\cite{Schulzc-BMJ2010-ConsortStatement,apa-jars,area-empirical-social-science-2006}.
\end{itemize}

\vspace{8pt}
\noindent
\textbf{Results}
\begin{itemize}
    \item Ensure that results directly address the research question(s)~\cite{Schulzc-BMJ2010-ConsortStatement,apa-jars,area-empirical-social-science-2006}.
    \item For qualitative results, provide tables, descriptions, quotes, and arguments to answer the research questions~\cite{apa-jars,area-empirical-social-science-2006}.
    \item For quantitative results, provide summary or descriptive statistics to answer the research questions~\cite{apa-jars,area-empirical-social-science-2006}. 
\end{itemize}

\vspace{8pt}
\noindent
\textbf{Threats to Validity}
\begin{itemize}
    \item Include a dedicated threats to validity section that lists internal, external, construct threats, and biases as appropriate for the research question(s) and study design~\cite{Schulzc-BMJ2010-ConsortStatement,apa-jars,area-empirical-social-science-2006}.
    \item Discuss how the study design addresses the threats.
    \item Discuss any threats not addressed by the study design.
    \item Explain and justify the unaddressed threats.
    \item Explain the potential impact on the interpretation of the study results due to the unaddressed threats~\cite{Schulzc-BMJ2010-ConsortStatement,apa-jars,area-empirical-social-science-2006}.
\end{itemize}

\subsection{Reviewer Guidelines for Reviewing Empirical Papers}

The responsibility for high-quality publications about CER work does not rest solely on authors. 
Review guidelines over the past several years have clarified expectations for reviewing empirical work, both as CER and experience reports. 
Reviewers have a responsibility to hold authors accountable for following norms for reporting CER. 
We recommend that reviewers use the author guidelines and standards documents above as a checklist for things to provide feedback on during peer review.  These guidelines and standards echo guidelines and standards from education and other fields about presentation of empirical work~\cite{Schulzc-BMJ2010-ConsortStatement,apa-jars,NSF19002,area-empirical-social-science-2006, WWC2020}.

\section{Threats to Validity}
\label{threats}
In developing our rubric and applying it to the work described in this paper, we identified some limitations and threats to validity we must address.  
We have backgrounds in conducting human-based empirical research in computing education, software engineering, and security, but do not explicitly have research degrees in education.  
However, two of authors have been members of the program committee of SIGCSE TS for multiple years, including years volunteering as program and symposium chairs, so we believe that we do have the needed background.

During the creation of the \textit{CER Empiricism Assessment Rubric}, we did not explicitly review or reference other validated rubrics.  
Therefore, it is possible that we omitted some aspects of educational research project design.  
To address this threat to validity, after we completed our work, we compared our rubric to other guidelines for reporting and assessing empirical research study quality, such as APA JARS~\cite{apa-jars}, the \emph{What Works Clearinghouse Standards Handbook} \cite{WWC2020} and CONSORT \cite{Schulzc-BMJ2010-ConsortStatement}, as discussed in Section \ref{reporting-quality}.  
We found that there was significant overlap in the core concepts of these other guidelines and rubrics with our own, indicating that our rubric captures many of the same reporting values.

In choosing the years and venues for inclusion in this review, we selected the conferences and journals that are widely considered to be the top tier for CER and only analyzed the proceedings or issues from two years, 2014 and 2015. 
We did not include more general education conferences that have a computing track, such as ASEE or FIE, and we excluded some other venues (e.g., Koli Calling) to make the review more feasible.
While this choice means we could have missed a portion of the community, we believe that the venues we chose cover the majority of the computing education literature.  
Further, some other venues accepted papers based solely on the abstract, rather than the entire paper, which we believe does not accurately represent the current state of empiricism.  
We chose the particular set of years to create a baseline on empirical work before more recent changes to paper tracks and reviewing guidelines at SIGCSE TS and other venues.
Also, we recognize that the findings from 2014 and 2015 may not represent the current state of empiricism in 2021.
Future work will consider a review of more recent publications.

Furthermore, we do not have the perspective of manuscripts that were not accepted into these venues.
As Fincher, et al., discussed in the conclusion of their book chapter~\cite{fincher_tenenberg_dorn_hundhausen_mccartney_murphy_2019}, a weakness of systematic literature reviews is that they focus on a quantitative analysis to characterize a body of work through an author defined lens and may miss the broader context of the timeframe in which the work was published. 
We do not consider the broader context of computing education during 2014 and 2015 and how that may impact the types of papers accepted.  
However, we minimize this limitation by focusing on characterizing empirical elements independent of the specific topic of the work (e.g., K-12 or CS1).  
Our work provides value by suggesting how the community can improve reporting standards independent of the broader community context for the acceptance and publication of the reviewed literature.

In our application of the rubric, we settled on a two-pass approach.  
Two researchers evaluated each paper using the \textit{Base Rubric} - evaluation method, evaluation subject, evaluation subject source, comparison, and number of participants.  
We used this approach to better ensure proper overall categorization of papers.  
We calculated the inter-rater reliability among the research team working on the categorization and determined only minor differences in the evaluations.  
After this initial evaluation, one researcher then completed the remaining items in the \textit{CER Empiricism Assessment Rubric} sub-rubrics as appropriate  for each paper.
We chose this approach because it gave us the best balance between efficiency and accuracy.  

However, we recognize the possibility that individual researchers mischaracterized aspects of individual papers. 
For example, one researcher might rate \textit{General Rubric: Research Objectives} for a paper as \textit{Partial and Not Labeled} because she found a discussion of research objectives, but not a dedicated section, while another reader could have missed the discussion altogether because he read too quickly.  We have provided our dataset~\cite{dataset_2021} for other researchers to consider when utilizing our rubric.

\section{Conclusion}
\label{conclusion}
Our research goal was to characterize the reporting of empiricism in Computing Education Research literature by identifying whether publications include content necessary for researchers to perform replications, meta-analyses, and theory building. 
This systematic literature review summarizes the type of papers and studies included during 2014 and 2015 in the SIGCSE TS, ICER, ITiCSE, TOCE, and CSE venues.  
A majority of the accepted papers report empirical work.
However those papers do not consistently follow reporting norms.
We have provided suggestions to authors and reviewers to move the community forward in publishing high-quality empirical work that can lead to meta-analysis and theory building.

We did observe progress in reporting empirical work in recent years. 
With the creation of the CS education research track at the SIGCSE TS for SIGCSE 2018~\cite{Perez-Quinones:2017:MPT:3129166.3129172}, the organizers updated the review criteria to specifically request reviewers evaluate the items we include in our \textit{General Rubric}. Additionally, TOCE now recommends that authors utilize APA JARS when organizing their submissions.

In future work, we will conduct a similar literature review on more recent publications to gauge whether the community has made any progress. With improved reporting, our next review will attempt to consider categorizing papers at a more granular level than survey, qualitative, and quantitative, which would provide the opportunity for more detailed rubrics.
We welcome feedback on our rubric for possible revisions in this future review.

As computing education is growing as a field and community, we need to establish norms for reporting empirical work.  
By doing so, we will support replication and meta-analysis. 
Increased rigor in reporting expectations will increase the reputation of CER in the broader computing research community, which will facilitate the growth and increased reputation of CER scholars and the computing education field.  
We all need to contribute: authors and researchers need to create and report well-designed, high-quality research studies or well-documented and supported experience reports; reviewers need to provide feedback not only on the novelty of the idea, but the quality of presentation; and the community needs to support replications and meta-analyses so we can grow our understanding of how to share computing education with the world.

\section{Acknowledgements}
\label{acks}
We would like to thank the students Brantley Collins and Lilian Scatalon who helped with applying the rubric and some data analysis. Additional thanks to the reviewers who provided excellent feedback on this paper and our plans for the next.
This material is based upon work supported by the National Science Foundation under Grants 1525373, 1525173, and 1525028.

\bibliographystyle{ACM-Reference-Format}
\bibliography{SLR}

\pagebreak
\section{Appendix A: CER Empiricism Assessment Rubric} 
\label{appendix-a}

\section*{Introduction}
The CER Empiricism Assessment Rubric was developed as a repeatable methodology for determining the degree and rigor in which empirical research principles are reported regarding research activity in CER literature.  The rubric makes no judgement on the quality of the research or of the research question of the reported project.  Instead, the rubric tries to establish the effectiveness of the paper at communicating the various aspects of the research activity to the reader such that it could not only be understood, but possibly replicated for further investigation.

\section*{Rubric Question Response Values}
For all sub-rubrics other than the Base Rubric, we only use values from the following scale, as discussed in Section 3.2:
\begin{itemize}
    \item Complete and Labeled
    \item Complete and Not labeled 
    \item Partial and Labeled
    \item Partial and Not Labeled
    \item Not Present
    \item Not Applicable
\end{itemize}

The descriptive values address two key two parts of reporting CER work - the completeness of the presentation of the information and whether the information is labeled in the paper or not.

\textit{Completeness -  the level of completeness of presented information}
\begin{itemize}
    \item Complete - Answers/addresses all questions for a rubric category.  There is no assessment on the quality of the answer.
    \item Partial - Answers/addresses some of the questions for a rubric category.  There is no assessment on the quality of the answer.
    \item Not Present - Answers/addresses none of the questions for a rubric category. The questions should  be addressed.
    \item Not Applicable - The rubric item is not applicable to the paper.
\end{itemize}

\textit{Labeled - whether the presented information is clearly labeled}
\begin{itemize}
    \item Labeled - There is a heading appropriate for the rubric item or there is emphasis (bold/italics) for the rubric item.
    \item Not Labeled - There is no heading or emphasis for the rubric item to easily find the item in the paper.
\end{itemize}

\section*{Rubric Steps}

Follow the steps highlighted in the boxes.  Key sub-activities associated with each step in the rubric are denoted by \S.

\begin{tcolorbox}[enhanced,drop shadow]
\textbf{Step 0: Read the research work you wish to evaluate.}
\end{tcolorbox}
\begin{tcolorbox}[enhanced,drop shadow]
\textbf{Step 1: Apply the Base Rubric to the research work.}
\end{tcolorbox}
\section*{BR) Base Rubric}
\label{base-rubric}
The Base Rubric consists of five high-level questions that provide basic information about a work, including the primary methodology, the subject being studied, where the subject originated, whether there is any comparison, and the size of the study as measured by the number of participants or data instances.

\subsection*{BR-1) Base Rubric: Evaluation Method}
After reading the paper, the first step is to determine the evaluation method used in the work.  
The evaluation method is a high-level categorization regarding the overall nature of the project that is being reported.
It is possible to select multiple options for the evaluation method, although any more than two would be highly unusual.  
An example of this could be a project that examined both student assessment scores and student survey responses when evaluating a new teaching method.  
In this case, the project could be categorized as \textit{Experimental, Survey}.

\subsubsection*{Literature Review}
A work is categorized as a literature review if it is mainly focused on reporting on the current state of the body of knowledge on a particular topic or research question.
The work does not necessarily have to add to the body of knowledge, but it often will draw conclusions on where the state of the research is heading.

\textit{\S : If the work is a literature review, complete item BR-2 (Evaluation Subject) from the Base Rubric (BR) only and then continue with Descriptive/Persuasive Rubric (DR).}

\subsubsection*{Exploratory}
An exploratory work is a research work-in-progress.  An exploratory project could originate from a model building exercise, observation without any predefined research questions, or building a framework or taxonomy.

\textit{\S : If the work is exploratory, complete the Base Rubric (BR) and the Experimental Rubric (ER).}

\subsubsection*{Descriptive/Persuasive}
A descriptive or persuasive work describes a current situation or paints a picture but does not test predictions nor does it imply any cause-and-effect relationships.  
This type of work is different than a literature review in that it is reporting on a research subject rather than the current body of knowledge.

\textit{\S : If the work is descriptive/persuasive, complete the Base Rubric (BR) and the Descriptive/Persuasive Rubric (DR).}

\subsubsection*{Survey}
Survey works use information gathered through forms or some other asynchronous querying method as their primary data source.
CER project often have surveys or similar instruments (such as student evaluations) as a secondary data source even when there are other data sources being used.

\textit{\S : If the work uses a survey as one of its evaluative methods, complete the Base Rubric (BR), the Experimental Rubric (ER), and the Survey Data Source Rubric (SR).}

\subsubsection*{Qualitative}
The primary indicator of an evaluative project with a qualitative evaluation method is the presence of free form or free text answers from participants.  
These responses then have to be individually coded or evaluated separately in order to draw any conclusions regarding the research questions of the project.
Qualitative data is common for projects with a small sample size, where in-person interviews or focus groups are used, or if participants are audio or video recorded in any way.  
In each of these cases, the data typically requires more time to evaluate, code, or process as opposed to quantitative data.

\textit{\S : If the work uses qualitative data as one of its evaluative methods, complete the Base Rubric (BR) and the Experimental Rubric (ER).}

\subsubsection*{Quantitative}
Whereas qualitative data is defined generally by free form information gathered from participants, quantitative data is identified by discrete values and counts that can be more easily and directly compared with each other.  
While it can be easier to identify a quantitative study that has defined control and treatment groups with a large amount of data, case studies, experience reports, and quasi-experimental studies can also be considered quantitative studies, depending on the types of data collected.
Quantitative data in CER is often gathered from Likert-type questions on participant surveys, course assessment data, enrollment and retention data, demographic data, and other information that could inform some quality or aspect of a pedagogical technique, course, or curricula.  

\textit{\S : If the work uses quantitative data as one of its evaluative methods, complete the Base Rubric (BR) and the Experimental Rubric (ER).}

\subsubsection*{Missing}
A work that states a causal relationship as fact but does not support that claim with any empirical research or data is considered to be missing the evaluation method.

\textit{\S : If the work's evaluation method is missing, there is no further categorization to be completed.}

\subsubsection*{Not Applicable}
Some CER venues publish works that require no evaluation, nor are taking any particular stance.  
These works could, for example, simply describe a new course or a new curriculum with no claims of effectiveness.  

\textit{\S : If the work does not require any empirical evaluation, there is no further categorization to be completed.}

\subsubsection*{Mixed Methods}
Any work that explicitly is using more than one method listed above is considered a mixed methods study.  

\textit{\S : If the work is considered a mixed methods study, denote each method in a comma-separated list and then complete all appropriate rubrics as listed with the method.}

\subsection*{BR-2) Base Rubric: Evaluation Subject}
The evaluation subject rubric item identifies the nature of the treatment that is being investigated by the work.  In a well-focused research work, there should only be one evaluation subject.  

\textit{\S : Select one of the following options as a part of the Base Rubric (BR).}

\subsubsection*{Pedagogical Technique}
A work is evaluating a pedagogical technique if the researchers are particularly interested in a specific teaching method.  Studies that are evaluating a pedagogical technique often focus on student learning outcomes as an indicator of the effectiveness of the treatment, which could be represented as assessment information or student perceptions of their learning in a course or for a particular knowledge unit.  These studies are also more likely to have threats to validity from instructor bias and should be addresses appropriately.

\subsubsection*{Tool}
A work is evaluating a tool if the research questions are addressing the effectiveness of the tool itself and not the underlying pedagogical technique.  This could also encompass various forms of educational technology, such as the effectiveness of distance learning tools.  While there is almost undoubtedly a pedagogical approach associated with a tool, selecting this option as the evaluation subject indicates that the implementation and usage of the tool itself is the primary focus.

\subsubsection*{Curriculum}
A curriculum work is a research project that is looking beyond just one particular course, but rather the creation and integration of multiple units or courses across a larger program.  This could also include special curricula, such as summer camps.

\subsubsection*{Assessment}
A research project where an assessment is the evaluation subject is focused on a particular assignment or quiz or set of assignments within a single knowledge unit or course.  These types of projects often look at types of assessments and what makes them effective, validating assessments, issues with scale, and academic integrity.

\subsubsection*{Community}
Community-based research projects attempt to ascertain some understanding of a group of individuals.  For examples, a project that is trying to determine the relative preparedness of underrepresented minority students before attending college could be categorized as community.

\subsubsection*{Other or Combination}
Other research works could fall outside the range of this rubric, or could be considered a combination of the items above.

\subsection*{BR-3) Base Rubric: Evaluation Subject Source}
The evaluation subject source rubric item establishes where the evaluation subject was first created.  In our examination of CER literature, we noticed that replication studies did not occur often or it was sometimes unclear if the treatment used was being introduced in the current paper of if this was a continuation of previous work.  Thus, this rubric item denotes whether the evaluation subject was created originally by the authors or elsewhere and whether the treatment was modified for use in the current study.

\textit{\S : Select one of the following options as a part of the Base Rubric (BR).}

\subsubsection*{Authors Here}
An evaluation subject is considered to be in this category if the authors created the treatment themselves for use in this study.

\subsubsection*{Authors Elsewhere}
If a subject was created by the authors elsewhere, then the treatment was first presented in an previous published work.

\subsubsection*{Other Modified}
A subject that was created by someone other than the authors and has been altered in some way would fit in this category.

\subsubsection*{Other Not Modified}
If the current work is a true replication study using a treatment that has not been altered and was created by another person, then this category is selected.

\subsubsection*{Community}
If the Evaluation Subject has been identified as \textit{Community}, then the Evaluation Subject Source will also be \textit{Community}, as it represents the idea that the subject is a self-identifying group of people related to an area of interest.

\subsection*{BR-4) Base Rubric: Comparison}
Sometimes a research question is intended to discover and report on the current state of the world.  However, many research questions aim to determine whether a treatment had an effect on a population.  If this is the case, the work should compare the results in some way to some form of control data.

\textit{\S : Select one of the following options as a part of the Base Rubric (BR).}

\subsubsection*{Historical}
If the results are compared to data from before the treatment, then indicate that the comparison is historical in nature.

\subsubsection*{Comparison}
If the results are compared to data generated as a part of the current research study, such as a from a specific control group or A/B testing, then select this category.

\subsubsection*{None}
If the results are reporting on the state of the world and are not compared to any other data set, indicate that there was no comparison.

\subsection*{BR-5) Base Rubric: Number of Participants}
The number of participants indicates the \textit{n} value of the study.  This could be the number of students in a course, the number of submissions to a grading system, the number of responses to a survey, etc.  

\textit{\S : Select one of the numeric range options as a part of the Base Rubric (BR).}

\begin{itemize}
    \item None
    \item 1
    \item 2
    \item 3-10
    \item 11-30
    \item 31-75
    \item 76-150
    \item 151-300
    \item 301-999
    \item 1000+
\end{itemize}

\begin{tcolorbox}[enhanced,drop shadow]
\textbf{Step 2: If the work was categorized as \textit{Exploratory}, \textit{Survey}, \textit{Qualitative}, or \textit{Quantitative}, complete the Experimental Rubric (ER).}
\end{tcolorbox}

\section*{ER) Experimental Rubric}
For each of these rubric items, answer using the guidelines listed in the Rubric Question Response Values section unless otherwise directed.  Each item has one or more questions that can aid you in determining the correct rubric value for the overall item.

\textit{\S : For each Experimental Rubric (ER) item, select one value from the Rubric Question Response Values that best describes how that particular item was reported in the paper unless otherwise directed.  Use the sub-questions with each item to help identify the proper selection.}

\subsection*{ER-1) Experimental Rubric: Research Objectives}
\begin{itemize}
    \item Does the paper include a description of the research objectives? (e.g., goals, questions, hypotheses)
\end{itemize}

\subsection*{ER-2) Experimental Rubric: Related Work}
\begin{itemize}
    \item Does the paper present related work?
    \item Does the paper link the research objectives directly to the related work?
\end{itemize}

\subsection*{ER-3) Experimental Rubric: Participants}
\begin{itemize}
    \item Does the paper provide demographics on the participants?
    \item Does the paper describe the sampling/recruitment methods used? (i.e., why these respondents? e.g., mailing list, advertised, etc)
\end{itemize}

\subsection*{ER-4) Experimental Rubric: Study Design}
\begin{itemize}
    \item Does the paper define the dependent and independent variables (including specific metrics)?
    \item Does the paper justify the variables in relevance to the overall research objective?
    \item Does the paper describe the treatments/protocol/steps followed in the study?
    \item For subjective measure, does the paper describe any type of inter-rater agreement?
\end{itemize}

\subsection*{ER-5) Experimental Rubric: Data Collection}
\begin{itemize}
    \item Does the paper describe who gathered the data?
    \item Does the paper describe how the data was gathered?
    \item Does the paper describe where the data was gathered?
\end{itemize}

\subsection*{ER-6) Experimental Rubric: Analysis Procedures}
\begin{itemize}
    \item Does the paper describe the analysis procedures? - process for working with the data after collection
    \item For Qualitative Data:
    \begin{itemize}
        \item Does the paper describe how they cleaned/coded the data?
        \item Does the paper describe how they evaluated the correctness of the coding (e.g. inter-rater reliability)?
    \end{itemize}
    \item For Quantitative Data:
    \begin{itemize}
        \item Does the paper describe the specific statistical tests that are used to analyze the data? (e.g., hypothesis checks, statistical tests, p-values, performance metrics, precision, recall, accuracy, False positive, False negative etc.)
        \item Does it justify why the tests were chosen?
    \end{itemize}
\end{itemize}

\subsection*{ER-7) Experimental Rubric: Results}
\begin{itemize}
    \item Does the paper include summary/descriptive statistics? (e.g., mean, std dev, charts/tables to describe data) 
    \item Does the paper discuss results in relation to the research objectives? (e.g., hypotheses evaluated, questions answered, or "big picture")
    \item Does the paper present the qualitative data, if applicable? (Tables, descriptions, arguments)
\end{itemize}

\subsection*{ER-8) Experimental Rubric: Threats to Validity}
\begin{itemize}
    \item Does the paper contain a dedicated discussion of the threats to validity (i.e., limitations or mitigations)?
    \item Does this section include a discussion of how the threats were addressed?
    \item Does this section include a discussion of threats left unaddressed?
\end{itemize}

\subsection*{ER-9) Experimental Rubric: Type of Study}
For this rubric item, determine whether the treatment is classified as observational or interventional in nature.

\textit{\S : Select one of the following options.}
\begin{itemize}
    \item Observational - Study is performed in a natural setting in which the researcher collects data via observation without manipulation of the situation. In this type of study, the researcher is merely observing the participants in a natural setting without interacting with the participants.
    \item Interventional - Study is performed by assigning participants into groups (e.g., control and experimental) and a treatment is applied to the experimental group to measure its effect. In this type of study, the researcher is interacting with the participants to study their response to certain variables.
\end{itemize}

\subsection*{ER-10) Experimental Rubric: Data Sources}
\textit{\S : Select all sources of data used in the study from this list.}

\begin{itemize}
    \item Survey
    \item Course Evaluations
    \item Assessment Data
    \item Observations
    \item Interviews
    \item Focus Groups
    \item Automated 
    \item Logs/Metadata/Generated/Mined
    \item Other
\end{itemize}

\begin{tcolorbox}[enhanced,drop shadow]
\textbf{Step 3: If the work was categorized as \textit{Survey}, complete the Survey Data Source Rubric (ER).}
\end{tcolorbox}

\section*{SR) Survey Data Source Rubric}
Complete this section if the evaluation method for the work was classified as survey or if one of the selected data sources for an experimental study was a survey.  The purpose of this rubric is to determine the quality of how information regarding the survey instrument was developed and presented.  

\textit{\S : For each Survey Data Source Rubric (SR) item, select one value from the Rubric Question Response Values that best describes how that particular item was reported in the paper.  Use the sub-questions with each item to help identify the proper selection.}

\subsection*{SR-1) Survey Data Source Rubric: Conducting the Survey}
\begin{itemize}
    \item Does the paper describe the study design? (e.g., pre- and post- surveys, reflections after assignments, etc.)
    \item Does the paper describe how the survey was administered? (e.g., in-person, remote)
    \item Does the paper describe the survey medium? 
\end{itemize}

\subsection*{SR-2) Survey Data Source Rubric: Survey Design}
\begin{itemize}
    \item Does the paper provide a rationale behind the questions? (i.e., why these questions and not others)
    \item Does the paper include the survey questions or provide a link to them?
\end{itemize}

\begin{tcolorbox}[enhanced,drop shadow]
\textbf{Step 4: If the work was categorized as \textit{Descriptive/Persuasive}, complete the Descriptive/Persuasive Rubric (DR).}
\end{tcolorbox}

\section*{DR) Descriptive/Persuasive Rubric}
Some papers that are published are not full research studies, but rather describe a current situation and do not imply any cause-and-effect relationships.  If this is the case, complete the following rubric items using the Rubric Question Response Values.

\textit{\S : For each Descriptive/Persuasive Rubric (DR) item, select one value from the Rubric Question Response Values that best describes how that particular item was reported in the paper.  Use the sub-questions with each item to help identify the proper selection.}

\subsection*{DR-1) Descriptive/Persuasive Rubric: Goal of the Argument}
\begin{itemize}
\item Does the paper describe the goal of the argument?
\end{itemize}

\subsection*{DR-2) Descriptive/Persuasive Rubric: Related Work (Context)}
\begin{itemize}
\item Does the paper present related work?
\item Does the paper link the research objectives directly to the argument?
\end{itemize}

\subsection*{DR-3) Descriptive/Persuasive Rubric: Premises and a Conclusion}
\begin{itemize}
\item Does the paper contain two or more premises and a conclusion? (Aristotle's rule)
\end{itemize}

\subsection*{DR-4) Descriptive/Persuasive Rubric: Supporting Evidence}
\begin{itemize}
\item Does the paper describe the supporting evidence?
\item Does the paper provide references for the the supporting evidence?
\end{itemize}

\end{document}